\documentclass[a4paper, 12pt]{article}

\usepackage{tikz-cd}
\usetikzlibrary{shapes.misc}
\usepackage{xcolor,tikz,pgfplots}
\usetikzlibrary{matrix,calc,positioning,decorations.markings,decorations.pathmorphing,decorations.pathreplacing}
\usetikzlibrary{arrows,cd}
\usepackage{bbding}
\usetikzlibrary{positioning}
\tikzset{every picture/.style={remember picture}}
\tikzset{cross/.style={cross out, draw=black, fill=none, minimum size=2*(#1-\pgflinewidth), inner sep=0pt, outer sep=0pt}, cross/.default={2pt}}
\usepackage{latexsym,amsmath,amsfonts,amssymb}
\usepackage{amsthm}
\usepackage{mathrsfs}
\usepackage[american]{babel}
\usepackage{bbm}
\usepackage{cite}    
\usepackage[pdfencoding=auto]{hyperref}
\hypersetup{colorlinks, citecolor=[rgb]{.7,0,0}, linkcolor=[rgb]{0,0,0.7}, urlcolor=[rgb]{0,0,0.5}}

\renewcommand{\baselinestretch}{1.2}
\numberwithin{equation}{section}

\newcommand{\be}{\begin{equation}} \newcommand{\ee}{\end{equation}}
\newcommand{\bea}{\begin{equation} \begin{aligned}} \newcommand{\eea}{\end{aligned} \end{equation}}
\newcommand{\ba}{\begin{array}} \newcommand{\ea}{\end{array}}

\newcommand{\cC}{\mathcal{C}}

\newcommand{\cH}{\mathcal{H}}

\newcommand{\bC}{\mathbb{C}}

\newcommand{\bZ}{\mathbb{Z}}

\makeatletter
\def\blfootnote{\gdef\@thefnmark{}\@footnotetext}
\makeatother

\begin{document}

\thispagestyle{empty}
\begin{flushright}
\end{flushright}
\vspace{13mm}  
\begin{center}
{\huge  On Continuous 2-Category Symmetries and Yang-Mills Theory\\[.5em] }

{\large Andrea Antinucci, Giovanni Galati, Giovanni Rizi}

\bigskip
{\it
SISSA, Via Bonomea 265, 34136 Trieste, Italy \\[.2em]
INFN, Sezione di Trieste, Via Valerio 2, 34127 Trieste, Italy \\[.2em]

}

\bigskip
{\rm
 aantinuc@sissa.it  \ \ 
 ggalati@sissa.it  \ \ 
 grizi@sissa.it 
}

\vspace{2cm}
\centerline{\bf Abstract}
\vspace{2 mm}
\begin{quote}
We study a 4d gauge theory $U(1)^{N-1}\rtimes S_N$ obtained from a $U(1)^{N-1}$ theory by gauging a 0-form symmetry $S_N$. We show that this theory has a global continuous 2-category symmetry, whose structure is particularly rich for $N>2$. This example allows us to draw a connection between the \emph{higher gauging} procedure and the difference between local and global fusion, which turns out to be a key feature of higher categorical symmetries. By studying the spectrum of local and extended operators, we find a mapping with gauge invariant operators of 4d $SU(N)$ Yang-Mills theory. The largest group-like subcategory of the non-invertible symmetries of our theory is a $\mathbb{Z}_N^{(1)}$ 1-form symmetry, acting on the Wilson lines in the same way as the center symmetry of Yang-Mills theory does. Supported by a path-integral argument, we propose that the $U(1)^{N-1}\rtimes S_N$ gauge theory has a relation with the ultraviolet limit of $SU(N)$ Yang-Mills theory in which all Gukov-Witten operators become topological, and form a continuous non-invertible 2-category symmetry, broken down to the center symmetry by the RG flow.

\end{quote}

\end{center}

\newpage
\pagenumbering{arabic}
\setcounter{page}{1}
\setcounter{footnote}{0}
\renewcommand{\thefootnote}{\arabic{footnote}}

{\renewcommand{\baselinestretch}{.88} \parskip=0pt
\setcounter{tocdepth}{2}
\tableofcontents}


\newpage
\section{Introduction}
The relevance of the concept of symmetry in quantum systems dates back to Wigner \cite{Wigner:1939cj}, who showed that a symmetry group $G$ is realized by (anti)linear and (anti)unitary operators $U_g$ on the Hilbert space, labeled by $g\in G$. In local quantum field theory global symmetry is the main tool. On the one side, it organizes the spectrum in representations of $G$, hinting which QFT can describe a given physical phenomenon. On the other side global symmetries and their anomalies are among the few intrinsic and renormalization group (RG) flow invariant properties \cite{tHooft:1979rat, Callan:1984sa}, imposing selection rules on correlation functions as well as constraints in strongly coupled theories. For instance, along with the RG flow, all the operators compatible with the global symmetries are generated by quantum effects, so that the full classification of the global symmetries of a model is a powerful tool to have control over the flow. This is the classic notion of \emph{naturalness} \cite{tHooft:1979rat}. It is very important to remark that even if a global symmetry is not exactly realized, it is often useful to study a limit in which the symmetry is restored and discuss its consequences. Then the amount of violation of these consequences will be estimated by the amount of violation of the symmetry. 

It is well known that except for few cases involving supersymmetry \cite{Seiberg:1994pq, Seiberg:1994bz, Seiberg:1994bp}, the standard global symmetries are not enough to constrain the RG flow and the infrared phase. Nevertheless, in some models there is evidence against the generation, along with the RG flow, of operators which do not violate the known symmetries of the theory \cite{Komargodski:2020mxz}. Our faith in the notion of naturalness generates a tension, which can be solved only by enlarging the \emph{category} of what we want to call global symmetries. A revolution took place in this sense in the last decade, starting from the observation \cite{Gaiotto:2014kfa} that global symmetries lead to extended unitary topological operators $U_g[\mathcal{M}_{d-1}]$ labeled by group elements $g\in G$, supported on co-dimension one manifolds, and following group-like fusion rules $U_g[\mathcal{M}_{d-1}]U_h[\mathcal{M}_{d-1}]=U_{gh}[\mathcal{M}_{d-1}]$. It has been noticed that it is really the topological nature of these operators which can replace the usual notion of symmetry, yielding by itself to their RG invariance, selection rules, anomalies, and the notion of naturalness. Quantum field theories however have much more topological operators that those supported on co-dimension one manifolds and following group-like fusion rules. Specifically, they can be supported in higher co-dimension manifolds, leading to the notion of higher $p$-form symmetries \cite{Kapustin:2014gua, Gaiotto:2014kfa} or they can be a mixture of higher form symmetries of different degrees $p$, producing the so-called higher groups \cite{Kapustin:2013uxa, Cordova:2018cvg, Benini:2018reh, Tachikawa:2017gyf} (see also \cite{Cordova:2020tij, Wan:2019soo, Hsin:2020nts, Apruzzi:2021mlh, DelZotto:2022joo, Bhardwaj:2022scy, Bhardwaj:2021wif}). But even more drastically there exist topological operators which are \emph{not} unitary, thus finding a way out from the Wigner paradigm, and do \emph{not} follow group-like fusion rules. Instead, the fusion of two defects produces a sum of several defects. In particular, there exist defects that do not possess an inverse under the fusion product, in sharp contrast with group-like symmetries. For this reason, this type of generalization is dubbed as \emph{non-invertible symmetries} \cite{Bhardwaj:2017xup, Chang:2018iay}.

In 2d theories, non-invertible symmetries generated by topological defect lines (TDLs) are ubiquitous \cite{Verlinde:1988sn, Frohlich:2006ch, Petkova:2000ip, Petkova:2013yoa}, and their correct underlying mathematical structure has been recognized to be that of certain 1-categories, namely \emph{fusion categories} \cite{etingof2005fusion, etingof2016tensor, Bhardwaj:2017xup, Chang:2018iay, Thorngren:2019iar, Thorngren:2021yso}. This means that the TDLs are objects of an Abelian category with a tensor product structure (fusion)
\begin{equation}
    L_a\otimes L_b=\sum _{c}f_{ab}^c L_c.
\end{equation}
The Abelian structure means that one can construct finite direct sums of objects, while the morphisms from $L_a$ to $L_b$ form a $\mathbb{C}-$vector space $\mbox{Hom}(a,b)$. Physically, the morphisms are topological local operators changing the line $L_a$ into $L_b$, and they can be combined linearly with arbitrary $\mathbb{C}-$coefficients. The structure constants $f_{ab}^c$ are the dimensions of $\mbox{Hom}(a\otimes b, c)$. Some lines cannot be written as a sum of others, and these are called \emph{indecomposable lines}. Importantly these lines are also \emph{simple}, meaning that they do not have endomorphisms on them, except for those proportional to the identity: $\mbox{Hom}(a,a)\cong \mathbb{C}$. Moreover if $a$ and $b$ are simple, there are no morphisms between them.\\
The dynamical consequences of having a fusion category symmetry have been studied, finding new constraints on the RG flow \cite{Chang:2018iay} and surprising solutions to problems of apparent lack of naturalness \cite{Komargodski:2020mxz}. These results demonstrate that exploring new examples of symmetries is not merely an academic exercise but can give interesting physical insights.\\ In 3d TQFT the line operators also form non-invertible symmetries with a similar categorical structure which takes into account the braiding, namely a \emph{modular tensor category} (MTC) \cite{Moore:1988qv, Moore:1989vd, Turaev:1994xb, Fuchs:2002cm, Kitaev:2005hzj, Kong:2013aya, Barkeshli:2014cna} (for a recent application in low dimensional holography see \cite{Benini:2022hzx}). Almost all the known examples are about discrete non-invertible symmetries, while continuous ones are believed to be rare and exotic. Moreover, until very recently, the existence of non-invertible symmetries in higher dimensions was seriously questioned, because of the lack of concrete examples. Last but not least, since non-invertible topological line defects are described by fusion category theory, it is natural to expect that the higher dimensional generalization is described by \emph{fusion higher category theory} \cite{douglas2018fusion, decoppet2021weak, decoppet2022multifusion}, which is a mathematical topic still in development. Roughly speaking, when one considers topological defects of dimension at least two, the morphisms are not local operators but extended ones. For instance, for a symmetry generated by surface defects, the morphisms are TDLs stacked on the defects, which are by themselves objects of a fusion category. They have their own morphisms, which in the higher category of surface operators are called \emph{2-morphisms}, and they form a vector space.

Despite the lack of a mathematically established structure, in the last year, many examples of non-invertible symmetries in higher dimensions have been discovered \cite{Nguyen:2021yld,Wang:2021vki,Choi:2021kmx, Kaidi:2021xfk, Roumpedakis:2022aik, Choi:2022zal, Bhardwaj:2022yxj, Kaidi:2022uux, Choi:2022jqy, Cordova:2022ieu,Hayashi:2022fkw,Arias-Tamargo:2022nlf}, which are related with some gauging procedure. In particular in \cite{Bhardwaj:2022yxj} many discrete 1-form non-invertible symmetries are constructed in 4d by generalizing a procedure well known in 3d TQFT \cite{Barkeshli:2014cna}, which consists in gauging a discrete invertible 0-form symmetry $G^{(0)}$ which acts as an automorphism of the set of generators $\left\{U_h\right\}_{h\in H}$ of a discrete 1-form symmetry $H^{(1)}$. The basic idea is the following. The generators of $H^{(1)}$ fall into various orbits $\mathcal{O}_{[h]}=\left\{ U_{g\cdot h} \; | \; g \in G \right\}$ for the action $g: \; h\rightarrow g\cdot h$ of $G^{(0)}$ on $H^{(1)}$. After gauging $G^{(0)}$ most of the $U_h$ are no longer gauge invariant. However, instead of throwing them away, we get new indecomposable objects labeled by the orbits $[h]$, each one being the sum of the objects in the corresponding orbit
\begin{equation}
\widehat{U}_{[h]}=\bigoplus \left\{U_{ h'}\; | \ [h']=[h] \right\}    
\end{equation}
up to a normalization factor. These new objects have \emph{not} group-like fusion rules. The orbits can be long or short, depending on the stabilizer. In the 3d procedure of \cite{Barkeshli:2014cna} the objects corresponding to short orbits come into copies labeled by representations of the stabilizer. This is because in $d$ dimensions the gauging of the 0-form symmetry produces a quantum $(d-2)-$form symmetry $\widehat{G}^{(d-2)}$, whose topological operators are the $G^{(0)}$ Wilson lines \cite{Gaiotto:2014kfa}. For $d=3$ this is a 1-form symmetry as the non-invertible one, and indeed they are both generated by lines. Therefore the line generators of the non-invertible symmetry can be \emph{dressed} with those of the quantum symmetry. Thus the full set of indecomposable objects in the gauged theory is given by $\widehat{U}_{[h]}^{\rho}=\widehat{U}_{[h]} \eta _{\rho}$, where $\rho \in \mbox{Rep}(G)$ and $\eta _{\rho}$ is the corresponding Wilson line. However, some of these $G^{(0)}$ Wilson lines are "absorbed" by the non-invertible lines, and one can argue that what effectively labels the different copies of $\widehat{U}_{[h]}$ are the irreducible representations of the stabilizer of $h\in H$ for the $G$ action on $H$ \cite{Barkeshli:2014cna}.

This \emph{dressing procedure} strictly speaking is no longer true in $d>3$ because the generators of the 1-form non-invertible symmetries have dimension $d-2>1$. The non-invertible symmetry is described by a $(d-2)-$category, and the Wilson lines of the dual symmetry appears as $(d-3)-$morphsims. In this paper, we will consider only the 4d case, in which the non-invertible 1-form symmetry is generated by topological surfaces and the quantum $\mbox{Rep}(G)$ symmetry is generated by lines, which can enter in the category of 1-morphisms of the surfaces. At first sight, the indecomposable objects of the gauged theory are labeled only by the orbits of the $G$-action on $H$. The approach of \cite{Bhardwaj:2022yxj} was that there are indeed no further objects, but the dual symmetry generated by the 1-endomorphisms should be sometimes gauged in the fusion rules. On this aspect we propose a somewhat different but equivalent point of view, which unifies the stories in 3d and higher dimensions. The idea is that the quantum 2-form symmetry $\mbox{Rep}(G)$, which is non-invertible for non-Abelian $G$, can be condensed on a surface $\Sigma$, generating the so-called \emph{condensation defects} \cite{Gaiotto:2019xmp, Roumpedakis:2022aik}. There are as many gauging procedures as many are the subgroups of $G$. In section \ref{sec:global fusion} we will give an alternative construction of the condensation defects which makes it clear that they are in one-to-one correspondence with the subgroups of $G$, also in the non-Abelian case. Our point of view is that in the gauged theory for each $G$-orbit we have many indecomposable defects, labeled by the subgroups of the stabilizer, and these are obtained from the \emph{naked} defect by dressing it with the various condensation defects. From this perspective, the 2-category must be enlarged by including all the dressed defects, similarly to the 3d case. This perspective is also motivated by the mathematical literature \cite{Gaiotto:2019xmp, douglas2018fusion}, which suggests that all the condensation defects must be added to obtain the so called \emph{idempotent completion}\footnote{Other names in the literature are Karoubi completion, or condensation completion.} of the fusion category.

This way of obtaining non-invertible symmetries by gauging automorphisms can produce a large number of examples \cite{Bhardwaj:2022yxj}, which is a very interesting "data-base" of higher category symmetries, potentially also for mathematicians. However, from a physical point of view, the theories one gets in this way are somewhat exotic. One of the aims of this paper is to provide an instance in which the gauging procedure is very natural, and is in some sense built-in. This is the case of the Weyl group $W_{G}$ in 4d $G$ Yang-Mills (YM) theory. If we denote by $N\left(U(1)^r\right)\subset G$ the normalizer of the Cartan torus $U(1)^r$ in $G$, the Weyl group is the quotient $W_G=N\left(U(1)^r\right)/U(1)^r$ and the normalizer can be written as $U(1)^r\rtimes W_G$\footnote{More precisely the normalizer fits in the short exact sequence
\begin{equation}
0\rightarrow U(1)^r \rightarrow N\left(U(1)^r\right)\rightarrow W_G \rightarrow 0 \end{equation}
defined by an action $\rho : W_G \rightarrow Aut\left(U(1)^r\right)$ and a non-trivial cocycle $e\in H_{\rho}^2\left(W_G, U(1)^r\right)$. As previously emphasized, we are mainly interested in the action of $\rho$ which specify how the zero-form symmetry acts on the generators of the one-form symmetry. Therefore the role of the cocycle $e$ in our discussion is marginal and we will neglect its effect in the following. It would be interesting to analyze its role in a future work.}. Therefore $W_G$ is automatically gauged in the $G$ YM theory. Does this produce a non-invertible symmetry? Strictly speaking the answer is no, basically because there is no theory producing $G$ YM theory upon gauging $W_G$. However, if we go to high energy where the theory becomes free\footnote{With an abuse of terminology by \emph{free} in the UV we will always mean \emph{weakly coupled}.}, a partial fixing of the non-Abelian gauge invariance is achieved by looking at the gauge theory for the Cartan torus $U(1)^r$ \cite{tHooft:1981bkw} (see section \ref{sec:3} for a detailed discussion). Here the  Weyl group appears as a global 0-form symmetry, and thus we need to gauge it to obtain a theory related to the UV limit of YM theory. We are led to look at the theory with gauge group given by the normalizer $U(1)^r\rtimes W_G$ of the Cartan torus. The Abelian gauge theory $U(1)^r$ has continuous 1-form symmetries on which the Weyl group acts as an automorphism\footnote{As pointed out in \cite{Brennan:2022tyl, Delmastro:2022pfo}, the interplay between the 1-form and the 0-form symmetry is fully specified only after we specify a symmetry fractionalization class in $H^2_{\rho}\left(W_G, (1)^r\right)$. This class is given by the cocycle $e$ which gives the extension $N\left(U(1)^r)\right)$ of $U(1)^r$ by $W_G$.}. Thus we are precisely in the condition described above, except that the 1-form symmetry is continuous. Then the $U(1)^r\rtimes W_G$ gauge theory is expected to have 1-form continuous non-invertible symmetries, described by a 2-category. We will focus on the case $G=SU(N)$, so we consider the $U(1)^{N-1}\rtimes S_N$ gauge theory in 4d. The 3d analog of this theory has been constructed on the lattice and with a different goal in \cite{Nguyen:2021yld}, where it has been dubbed \emph{semi-Abelian theory}. In that paper it is also pointed out that there are non-invertible symmetries. However, their fusion rules have not been computed, and only a subset of these symmetries has been discussed. In particular, even if it is pointed out that the general topological operators are parametrized by $N-1$ parameters, the ones studied in \cite{Nguyen:2021yld} depends only on one compact variable. On the other hand, we will see that the parameter space of the non-invertible symmetry is $U(1)^{N-1}/S_N$, and that the fusion rules are
\begin{equation}
      \mathcal{T}(\boldsymbol{\alpha})\otimes \mathcal{T}(\boldsymbol{\beta})=\sum _{\sigma \in H_{\boldsymbol{\alpha} } \backslash S_N /H_{\boldsymbol{\beta}}}f_{\alpha \beta}^{\sigma}\;\; \mathcal{T}\left(\boldsymbol{\alpha}+\mathfrak{S}^{\vee}_{\sigma }\cdot \boldsymbol{\beta} \right)
\end{equation}
where
\begin{equation}
    f_{\alpha \beta}^{\sigma} = \frac{|H_{\boldsymbol{\alpha}+\mathfrak{S}^{\vee}_{\sigma }\cdot \boldsymbol{\beta}}|}{|H_{\boldsymbol{\alpha}}\cap \sigma H_{\boldsymbol{\beta}} \sigma ^{-1}|} .
\end{equation}
Here $\mathfrak{S}_{\sigma}^{\vee}$ is the relevant action of the permutation group $S_N$ on the labels $\boldsymbol{\alpha}\in U(1)^{N-1}$, while $H_{\boldsymbol{\alpha}}\subset S_N$ denotes the stabilizer for this action. One of our main results is that, while the fusions above hold when the defect does not contain non-trivial 1-cycles, on a general topology we have to modify the formula above by including the condensations
\begin{equation}
\mathcal{T}(\boldsymbol{\alpha})[\Sigma]\otimes  \mathcal{T}(\boldsymbol{\beta})[\Sigma]   =\sum _{\sigma \in H_{\boldsymbol{\alpha}}\backslash S_N /H_{\boldsymbol{\beta}}}f_{\alpha \beta}^{\sigma} \; P_{\mbox{Rep}\left(\left(H_{\boldsymbol{\alpha}}\cap H_{\boldsymbol{\beta}}\right)^{\perp}\right)}\otimes \mathcal{T}\left(\boldsymbol{\alpha} +\mathfrak{S}_{\sigma} ^{\vee} \cdot \boldsymbol{\beta}\right)[\Sigma]\;.
\end{equation}
The operator $P_{\mbox{Rep}\left(\left(H_{\boldsymbol{\alpha}}\cap H_{\boldsymbol{\beta}}\right)^{\perp}\right)}$ coincide, up to a normalization, with a condensation defect, and it is a projector in the sense that it squares to itself.

Notice that $U(1)^{N-1}/S_N$ coincides with the set of conjugacy classes of $SU(N)$ also labeling the Gukov-Witten (GW) operators of $SU(N)$ YM theory \cite{Gukov:2006jk, Gukov:2008sn, Gukov:2014gja}. In the full YM theory, only the GW labeled by central elements are topological, and generate the 1-form center symmetry. We propose that \emph{all} the GW operators in YM theory become topological at high energy and form a non-invertible symmetry, broken down to the center symmetry by the RG flow. That the $SU(N)$ YM theory at high energy has non-invertible symmetries has been recently observed from a different point of view also in \cite{Cordova:2022rer}. The fact that these two distinct arguments agree is reassuring. Moreover, in that paper, the fusion rules have been computed only in the $N=2$ case, and they agree with those of the $U(1)\rtimes S_2$ gauge theory\footnote{The coefficients appearing are different, but this has to do with different choices of normalization. However, as we will see, with our normalization the fusion coefficients turn out to be always integers and as we point out in the main text, this is important since they count the number of 1-morphisms up to endomorphisms.}. It is reasonable that by applying the methods of \cite{Cordova:2022rer} for any $N$ one gets the fusion rules which we compute in the $U(1)^{N-1}\rtimes S_N$ theory, thus confirming that the symmetry found in that paper is really the same discussed here. We leave this interesting problem for future work. In 2d YM theories it was already pointed out in \cite{Nguyen:2021naa} that all the GW operators are topological, and they form a non-invertible symmetry at all energy scales. This conclusion is peculiar of 2d YM theory since the theory is quasi-topological. In $d>2$ this is obviously not true and indeed this symmetry exists only in the UV limit.

The connection between the UV symmetries of YM theory and those of the $U(1)^{N-1}\rtimes S_N$ gauge theory is important because the second case is much more under control, and we are able to discuss the 2-categorical structure of this symmetry (section \ref{sec:global fusion}), which indeed was not analyzed before. The analysis of this structure is the bulk of the paper. In our examples, we find several properties which we believed to be general aspects of 2-category symmetries. For instance, we argue that the almost universal presence of condensation defects on the right-hand side of the fusion rules is what distinguishes the global fusion rules (those obtained on general manifolds) from the local ones, which are true only if the defects are topologically trivial. Moreover, we find that the fusion coefficients are always positive integer numbers. We interpret these numbers as counting the 1-morphisms up to possible endomorphisms. This is an important difference with respect to fusion 1-category symmetries, in which the indecomposable objects cannot have non-trivial endomorphisms, and therefore the fusion coefficients are directly counting the morphisms living at the junctions. Moreover, while in fusion 1-categories the morphisms form a vector space, and therefore these numbers are the dimensions of these vector spaces, in fusion 2-categories the 1-morphsims form by itself a category, and the numbers are better interpreted as quantum dimensions. Finally, after understanding the map between the gauge invariant operators of YM theory and the $U(1)^{N-1}\rtimes S_N$ gauge theory, we are able to determine how this non-invertible symmetry acts on line operators which are compatible with the known results when we restrict to the group-like subcategory $\mathbb{Z}_N^{(1)}$ corresponding to the center.

The rest of the paper is organized as follows. In section \ref{sec:2} we study the  $U(1)^{N-1}\rtimes S_N$ gauge theory, by analyzing the full spectrum of gauge-invariant operators, finding the continuous non-invertible 1-form symmetries. The intricate 2-categorical structure of this symmetry is analyzed in section \ref{sec:global fusion}, where we also explain the connection, in our specific example, between the concept of \emph{global fusions} introduced in \cite{Bhardwaj:2022yxj} and the \emph{higher condensation defects} constructed in \cite{Roumpedakis:2022aik}. Then section \ref{sec:3} is devoted to the connection between the $U(1)^{N-1}\rtimes S_N$ gauge theory and $SU(N)$ YM theory at high energy. After a path integral argument, we show a mapping among all kinds of gauge invariant observables of the two theories (local, line, and surface operators). Then we identify the center symmetry of $SU(N)$ YM theory with a discrete subset of topological surface operators of the $U(1)^{N-1}\rtimes S_N$ theory, by showing that they give rise to the same Ward identities with the Wilson line operators. We also discuss how all the possible choices of the global structure of the YM theory are obtained from the point of view of the $U(1)^{N-1}\rtimes S_N$ gauge theory. We conclude in section \ref{sec:4} with a discussion on possible future directions.

\section{The 4d $U(1)^{N-1}\rtimes S_N$ Gauge Theory}
\label{sec:2}
This section is devoted to the $U(1)^{N-1}\rtimes S_N$ gauge theory on its own. We show that the theory has non-invertible 1-form symmetries labeled by continuous parameters valued in $U(1)^{N-1}/S_N$. This non-invertible symmetry is described by a 2-category which we study in detail, discovering an interesting mathematical structure.
\subsection{Abelian Gauge Theory}\label{sec:Abeliangaugetheory}
We start with a free Abelian gauge theory with gauge group $U(1)^{N-1}$. The definition of the theory is encoded in the choice of the spectrum of Wilson line operators, namely an $N-1$ dimensional lattice. A way to make this explicit is by exhibiting a basis for the gauge fields $\mathcal{A}_{i=1,...,N-1}$ in which the Wilson lines have integer charges. This is a choice of a symmetric positive definite $(N-1)\times (N-1)$ matrix $Q^{(N-1)}_{ij}$ such that the action is
\begin{equation}
    S=\frac{1}{2e^2}\int d^4 x \;Q^{(N-1)}_{ij}\mathcal{F}_i\wedge * \mathcal{F}_j
\end{equation}
where $\mathcal{F}_i=\ d \mathcal{A}_i$. Then the most general Wilson line is 
\begin{equation}
    \label{eq: general Abelian Wline}
   \mathcal{W}(\boldsymbol{n})[\gamma]= \mathcal{W}(n_1,...,n_{N-1})[\gamma]=\prod _{i=1}^{N-1}\mathcal{W}_i[\gamma] ^{n_i} \ , \ \ \ \ \mathcal{W}_i [\gamma] ^{n_i}:=\exp{\left(in_i\oint _{\gamma} \mathcal{A}_i\right)}
\end{equation}
where $\boldsymbol{n}=(n_1,...,n_{N-1})\in \mathbb{Z}^{N-1}
\label{eq:globalchoice}$. To make the action of the $S_N$ 0-form symmetry explicit we define the theory by demanding that, upon introducing $\mathcal{A}_N=-\mathcal{A}_1-...-\mathcal{A}_{N-1}$ the action takes the form\footnote{Here $\mathcal{F}_i\mathcal{F}_j$ means $\mathcal{F}_i\wedge *\mathcal{F}_j$.}
\begin{equation}
S=\frac{1}{2e^2}\int d^4 x\left(\mathcal{F}_1^2+...\mathcal{F}_N^2\right)= \frac{1}{2e^2}\int d^4 x\left(\sum _{i=1}^{N-1}2\mathcal{F}_i^2+\sum _{i<j}^{N-1}2\mathcal{F}_i\mathcal{F}_j\right)=\frac{1}{2e^2}\int d^4 x Q^{(N-1)}_{ij}\mathcal{F}_i\mathcal{F}_j
\end{equation}
thus defining the quadratic form $Q^{(N-1)}$ as
\begin{equation}
    Q^{(N-1)}_{ij}=1+\delta _{ij} \ , \  \mbox{with} \ \ \ \ \  \text{det}(Q^{(N-1)})=N \ , \ \ \  \left(Q^{(N-1)}\right)^{-1}_{ij}=\frac{-1+N\delta _{ij}}{N}.
    \label{eq:definitionQ}
\end{equation}
 
The $S_N$ symmetry permutes the connections $\mathcal{A}_{i=1,...,N}$. On the $N-1$ field strengths $\mathcal{F}_1,...\mathcal{F}_{N-1}$ it acts in the \emph{standard representation} of $S_N$, which we denote by $\mathfrak{S}$ (see appendix \ref{sec:SN}). This obviously induces also an action of $S_N$ on the Wilson lines, which is conveniently rewritten as an action on the charges:
\begin{equation}
\label{eq:action S_n on Wlines}
   \sigma \cdot \mathcal{W}(\boldsymbol{n})=\exp{\left(i\oint\sum _{j=1} ^{N-1}n_j\mathfrak{S}_{\sigma}\left(\mathcal{A}_j\right)\right)}=\exp{\left(i\oint\sum _{j=1} ^{N-1}\mathfrak{S}_{\sigma ^{-1}}^{\vee}( n_j)\mathcal{A}_j\right)}=\mathcal{W}\left(\mathfrak{S}_{\sigma ^{-1}}^{\vee}\cdot \boldsymbol{n}\right) \; .
\end{equation}
We have introduced $\mathfrak{S}_{\sigma} ^{\vee}\cdot \boldsymbol{n}=\left(\mathfrak{S}_{\sigma}^{\vee}(n_1),...,\mathfrak{S}^{\vee}_{\sigma}(n_{N-1})\right)$, with $\mathfrak{S}^{\vee}_{\sigma}(n _i)=m_{\sigma(i)}-m_{\sigma(N)}$, where $n_i=m_i-m_N$. This is a dual representation of $S_N$ on $N-1$ variables (see appendix \ref{sec:SN}). 

We also have electric GW operators \cite{Gukov:2006jk, Gukov:2008sn} (for a review \cite{Gukov:2014gja})
\begin{equation}
    \label{eq: general abelin eGW}
    \mathcal{D} (\boldsymbol{\alpha})[\Sigma]=\mathcal{D}(\alpha _1,...,\alpha _{N-1})[\Sigma ]=\prod _{i=1}^{N-1} \mathcal{D}_i (\alpha _i)[\Sigma] \ , \  \ \ \mathcal{D}_i(\alpha_i)[\Sigma]:=\exp{\left(i\alpha _i \int _{\Sigma} \frac{*\mathcal{F}_i}{e^2}\right)}.
\end{equation}
The variables $\boldsymbol{\alpha}=(\alpha _1,...,\alpha _{N-1})$ parametrize an $(N-1)$-dimensional torus (the precise periodicity is shown below). These operators are the generators of the electric 1-form symmetry \cite{Gaiotto:2014kfa}  $\left(U(1)_{e}^{(1)}\right)^{N-1}$.
On the GW operators $S_N$ acts as it does on the Wilson lines:
\begin{equation}
\sigma \cdot \mathcal{D}(\boldsymbol{\alpha})= \mathcal{D}\left(\mathfrak{S}_{\sigma ^{-1}}^{\vee} \cdot \boldsymbol{\alpha}\right).
\end{equation}
The electric GW operators $\mathcal{D}(\boldsymbol{\alpha})$ have an action on the Wilson lines $\mathcal{W}(\boldsymbol{n})$ by linking, and a simple computation shows the following Ward identity
\begin{equation}
\label{eq:GW on Lines Abelian}
\mathcal{D}(\boldsymbol{\alpha})[\Sigma ]   \cdot \mathcal{W}(\boldsymbol{n})[\gamma]=\exp{\left(i\;  Lk(\Sigma ,\gamma)\sum _{i,j=1} ^{N-1}\alpha _i \left(Q^{(N-1)}\right)^{-1}_{ij} n_j \right)} \mathcal{W}(\boldsymbol{n})[\gamma]
\end{equation}
where $ Lk(\Sigma ,\gamma)$ denotes the linking number between $\Sigma$ and $\gamma$. From this we deduce the periodicity $\alpha _i \sim \alpha _i +2\pi w_jQ_{ji}$, $w_i\in \mathbb{Z}$. Equivalently, the variables 
$$
\beta _i:=\alpha _i \left(Q^{N-1}\right)^{-1}_{ji}
$$
are $2\pi$ periodic, thus parametrizing a torus $U(1)^{N-1}$.

An analogous discussion holds for the \emph{'t Hooft lines} $\widetilde{\mathcal{W}}(\boldsymbol{n})$ and magnetic GW operators $\widetilde{\mathcal{D}}(\boldsymbol{\alpha})$. However the global structure we have chosen restricts the set of allowed 't Hooft lines by Dirac quantization conditions. We will discuss this in detail in section \ref{sec:global_structures}.

\subsection{Warm Up: $N=2$ Case}
\label{sec:N=2}
Before we face the general case, it is useful to study the baby example $N=2$, which is simpler since $S_2=\mathbb{Z}_2$ is Abelian, but captures several features of the general case. Indeed $U(1)\rtimes \mathbb{Z}_2=O(2)$ and the model is known as the $O(2)$ gauge theory \cite{Kiskis:1978ed,Schwarz:1982ec}. This subsection contains a review of discussions in \cite{Heidenreich:2021xpr} and \cite{Bhardwaj:2022yxj}, where the model has been shown to have non-invertible symmetries, but we also introduce new points, which we will expand on in the general case.

We start from the $U(1)$ Maxwell theory, in which $S_2=\mathbb{Z}_2$ acts as charge conjugation by reversing the sign of the connection $\mathcal{A}$, we then gauge this symmetry obtaining the $U(1)\rtimes \mathbb{Z}_2$ theory. A class of operators of this theory consists in gauge invariant operators of the $U(1)$ theory. but we also have the $\mathbb{Z}_2$ Wilson line
\begin{equation}
    \eta [\gamma]= e^{i\oint_{\gamma} a_2}
\end{equation}
where $a_2 \in H^1(\mathcal{M}_4,\mathbb{Z}_2)$ is the dynamical $\mathbb{Z}_2$ gauge field. The $\eta$ line is topological and generates the quantum 2-form symmetry $\widehat{\mathbb{Z}}_2^{(2)}$ as $\eta^2 = 1$. 

 Let us discuss the $\mathbb{Z}_2$ invariant combinations of operators of the original Abelian theory, which remain good operators after gauging. The local operators are all the even polynomials in the field strength. The Wilson lines $\mathcal{W}(n)$ of the Maxwell theory are labeled by one integer, their charge, and $\mathbb{Z}_2$ acts by reversing the sign of $n$. The Wilson line operators of the $O(2)$ gauge theory are obtained from those of $U(1)$ by summing over the $\mathbb{Z}_2$ orbits:
\begin{equation}
    \mathcal{V}(n)[\gamma]=\mathcal{W}(n)[\gamma]+\mathcal{W}(-n)[\gamma]=e^{in\oint _{\gamma} \mathcal{A}}+e^{-in\oint _{\gamma} \mathcal{A}}.
\end{equation}
We have a similar story for the electric GW operators. Imitating the well-known 3d procedure of \cite{Barkeshli:2014cna} described in the introduction, we build the gauge-invariant surface operators by summing over the $\mathbb{Z}_2$ orbits. For reasons that will be clear in the following, we normalize the operators by dividing them by $|H_{\alpha}|$, where $H_{\alpha}\subset \mathbb{Z}_2$ is the stabilizer of $\alpha$ 
\begin{equation}
\mathcal{T}(\alpha)[\Sigma]=\frac{1}{|H_{\alpha}|}\left(\mathcal{D}(\alpha)[\Sigma]+\mathcal{D}(-\alpha)[\Sigma]\right)= \frac{1}{|H_{\alpha}|}\left(e^{i\alpha \int _{\Sigma} \frac{*\mathcal{F}}{e^2}}+e^{-i\alpha \int _{\Sigma} \frac{*\mathcal{F}}{e^2}}\right). \end{equation}
In this case $H_{\alpha}$ can be either trivial (for $\alpha \neq 0,2\pi$) or equal to $\mathbb{Z}_2$ (for $\alpha =0,2\pi$). The operators $\mathcal{T}(\alpha)$ are \emph{indecomposable} objects after gauging, meaning that they cannot be written as direct sum of other objects. Note that with this normalization  we always define the operators $\mathcal{T}(\alpha)[\Sigma]$ as the direct sum of $\mathcal{D}$ without any coefficient. In particular
\begin{equation}
    \begin{split}
        &\mathcal{T}(\alpha)[\Sigma] = \mathcal{D}(\alpha)[\Sigma] + \mathcal{D}(-\alpha)[\Sigma]\qquad \alpha \not= 0,\pi\\
        &\mathcal{T}(\alpha)[\Sigma] = \mathcal{D}(\alpha)[\Sigma] \qquad \alpha = 0,\pi.
    \end{split}
\end{equation}
With other normalizations we either get fractional coefficients (which are meaningless in a categorical language) or \emph{decomposable} objects. This is the very reason for our choice of normalization, which we will keep also in the general $N$ case.

Since $Q^{(1)}=2$ in our normalization, $\mathcal{D}(\alpha)$ is parametrized by $\alpha \in [0,4\pi)$. Then the manifold where $\alpha$ takes values in the $O(2)$ theory is $U(1)/\mathbb{Z}_2=[0,2\pi]$, which is singular since $\alpha =0,2\pi$ are fixed points of the $\mathbb{Z}_2$ action. The somewhat surprising fact is that, since these operators are topological, they can be regarded as the generator of a symmetry, even though $\mathcal{T}(\alpha)$ is not a unitary operator and does not satisfy a group law multiplication:
\begin{equation}
\label{eq:local fusion O(2)}
    \mathcal{T}(\alpha)\otimes \mathcal{T}(\beta)=\frac{1}{|H_{\alpha}||H_{\beta}|}\Big(|H_{\alpha+\beta}|\mathcal{T}(\alpha+\beta)+|H_{\alpha-\beta}|\mathcal{T}(\alpha-\beta)\Big).
\end{equation}
This is a non-invertible symmetry \cite{Bhardwaj:2017xup}.  In the last few years these new type of symmetries have been analyzed extensively in 2d (for instance \cite{Bhardwaj:2017xup,Chang:2018iay,Thorngren:2019iar,Komargodski:2020mxz,Thorngren:2021yso}), and very recently also in higher dimensions
\cite{Choi:2021kmx, Kaidi:2021xfk, Roumpedakis:2022aik, Choi:2022zal, Bhardwaj:2022yxj, Choi:2022jqy, Cordova:2022ieu}. However most of the examples in the literature discuss \emph{discrete} non-invertible symmetries, while the non- invertible symmetry of the $O(2)$ gauge theory, as well as the other cases we discuss in the present paper are \emph{continuous} non-invertible symmetries. Until recently, these where believed to be very rare and exotic type of symmetries. One of our aims is to show that they can appear quite naturally, and they have some features similar to more common continuous symmetries. 

Notice that there are exactly two values of $\alpha$ for which the fusion is group-like, namely the fixed points of the $\mathbb{Z}_2$ action $\alpha =0,2\pi$, for which
\begin{equation}
    \mathcal{T}(2\pi)\otimes \mathcal{T}(2\pi)=\mathcal{T}(4\pi)=\mathcal{T}(0)=1.
\end{equation}
These are also the only two unitary operators. This shows that the large and continuous non-invertible symmetry contains an invertible $\mathbb{Z}_2^{(1)}$ 1-form symmetry, which is nothing but the center symmetry since $\mathcal{Z}(O(2))=\mathbb{Z}_2$. It is also important to notice that in the fusion \eqref{eq:local fusion O(2)} the coefficients are always integer numbers. This is obvious when $H_{\alpha}$ and $H_{\beta}$ are both either trivial or $\mathbb{Z}_2$. When instead $H_{\alpha}=1$ but $H_{\beta}=\mathbb{Z}_2$ the $1/2$ factor is cancelled because $\mathcal{T}(\alpha+\beta)=\mathcal{T}(\alpha-\beta)$. We will show that an analogous mechanism takes place for general $N$. This fact is important because the fusion coefficients have a meaning and must be integer numbers: when $\mathcal{T}(\gamma)$ appears in the fusion $\mathcal{T}(\alpha)\otimes \mathcal{T}(\beta)$ it means that there is a fusion category of 1-morphisms $\mathcal{T}(\alpha)\otimes \mathcal{T}(\beta)\rightarrow \mathcal{T}(\gamma)$, and the coefficient counts the number of simple lines in this category, or more precisely its \emph{total quantum dimension}. However, since some objects have non-trivial endomorphisms, this counting is only up to these endomorphisms. We will expand on this point in the general case.

The non-unitarity of the GW operators $\mathcal{T}(\alpha)$ for $\alpha \neq 0, 2\pi$ reflects itself in the fact that the charges of Wilson lines are not phases, as follows from the generalized Ward identity
\begin{equation}
\mathcal{T}(\alpha)[\Sigma]\cdot \mathcal{V}(n)[\gamma]=\frac{2}{|H_{\boldsymbol{\alpha}}|}\cos{\left( Lk(\Sigma,\gamma)\; n\frac{\alpha}{2}\right)} \mathcal{V}(n)[\gamma].
\end{equation}
We get a phase only for $\alpha =0,2\pi$ in which the GW operators are group-like. This phase is $(-1)^n$ depending only on the parity of $n$. Notice that at generic values of $\alpha$, different $n$'s with the same parity give different charges.

Up to this point, the discussion was a bit naive and indeed was correct only in the case when $\Sigma$ does not have non-trivial 1-cycles \cite{Bhardwaj:2022yxj}. When we consider topologically non-trivial defects, we need to modify the discussion above and analyze in detail the 2-categorical structure of the non-invertible symmetries. To do this, we have to incorporate the dual 2-form symmetry $\mathbb{Z}_2^{(2)}$ arising from the gauging. This story will be more complicated in the general case $N>2$ in which $S_N$ is non-Abelian, so it is worth discussing the symmetry structure before in this simple example. Before gauging, the electric 1-form symmetry has a very simple 2-categorical structure: the indecomposable objects are $\left\{\mathcal{D}(\alpha)\right\}_{\alpha \in [0,4\pi)}$ and the category of 1-morphisms $\mathcal{D}(\alpha)\rightarrow \mathcal{D}(\beta)$ is empty unless $\alpha=\beta$, in which case it contains only the identity line. After gauging, we get one additional topological operator, namely the non-trivial $\mathbb{Z}_2$ Wilson line $\eta$, which does not affect the indecomposable objects but enters in the 1-morphisms. This is a sharp difference with respect to the 3d case of \cite{Barkeshli:2014cna} in which by \emph{dressing} the objects with $\eta$ one gets new indecomposable objects. Naively, in 4d it seems that there are no further indecomposable objects, but we will explain shortly that this conclusion is wrong.

Since $\eta$ is a bulk line, it exists as a 1-morphism $\eta : \mathcal{T}(0)\rightarrow \mathcal{T}(0)$, but also as a 1-morphism on the surface $\mathcal{T}(2\pi)$ on which it is non-trivial\footnote{This is because the operators $\mathcal{T}(0),\mathcal{T}(2\pi)$ were indecomposable objects also in the pre-gauged theory, and they do not see the $\mathbb{Z}_2$ symmetry. Therefore it is not required to put boundary conditions for the $\mathbb{Z}_2$ on the gauge field.}. Notice an important difference of higher category symmetries with respect to more standard fusion categories of topological defect lines in 2d \cite{Bhardwaj:2017xup,Thorngren:2019iar}: even for indecomposable objects, the category of 1-endomorphism can contain non-trivial operators because there can be lower dimensional topological bulk defects which can be put on the objects without becoming trivial. As we will see, there are further interesting cases in which additional topological lines exist only stacked on a non-trivial surface. The surface operators $\mathcal{T}(\alpha)$, $\alpha\neq 0,2\pi$ on the other hand absorb the Wilson line $\eta$. Therefore the only 1-endomorphism on them is the identity. This is because before gauging $\mathcal{D}(\alpha)$, $\alpha\neq 0,2\pi$ is not invariant under $\mathbb{Z}_2$, so the precise definition of the gauge invariant defect $\mathcal{T}(\alpha)=\mathcal{D}(\alpha)+\mathcal{D}(-\alpha)$ requires to fix Dirichlet boundary conditions for the $\mathbb{Z}_2$ gauge field on the surface. We will call these kinds of objects \emph{strongly simple}, following the terminology of \cite{Johnson-Freyd:2020ivj}.

The discussion above is crucial whenever $\Sigma$ has non-contractible 1-cycles. When this is the case, the same line $\eta$ can be non-trivial on the surface, and generates a 0-form symmetry $\mathbb{Z}_2$ on it. As suggested in \cite{Bhardwaj:2022yxj}, the \emph{local} fusion rules \eqref{eq:local fusion O(2)} must be modified by generally gauging this 0-form symmetry on $\Sigma$, leading to the \emph{global} fusion rules. We understand this gauging procedure as well as the necessary modification of the fusion by a different argument. One can use the 2-form symmetry $\mathbb{Z}_2^{(2)}$ in the bulk to construct one further topological surface operator by condensing the symmetry on a surface, as explained in detail in \cite{Roumpedakis:2022aik}:
\begin{equation}
    \label{eq:condensation defect Z2}
    \mathcal{C}[\Sigma]:=\frac{1}{\sqrt{|H_1\left(\Sigma,\mathbb{Z}_2\right)|}}\sum _{\gamma \in H_1\left(\Sigma,\mathbb{Z}_2\right)} \eta [\gamma].
\end{equation}
Even if it is a surface operator, it has trivial action on lines because it is made of lower dimensional objects which cannot braid with lines. Notice that the condensation produces a dual 0-form $\mathbb{Z}_2$ symmetry living on the defect, which is generated by topological lines. The condensation  defect is non-invertible, and its fusion was computed in \cite{Roumpedakis:2022aik} to be
\begin{equation}
    \label{eq:fusion condensation Z2}
    \mathcal{C}[\Sigma]\otimes\mathcal{C}[\Sigma]=\mathcal{Z}(\mathbb{Z}_2;\Sigma) \mathcal{C}[\Sigma]
\end{equation}
where $\mathcal{Z}(\mathbb{Z}_2;\Sigma)=\sqrt{|H_1\left(\Sigma,\mathbb{Z}_2\right)|}$ is the partition function of the 2d pure $\mathbb{Z}_2$ gauge theory on $\Sigma$. The fact that the fusion coefficients are not numbers, but partition functions of TQFT, seems to be a general feature of higher category symmetries, as pointed out in recent papers \cite{Roumpedakis:2022aik,Choi:2022zal}. We will derive the same result from a different point of view in subsection \ref{sec:global fusion}, also generalizing to the case in which the symmetry that we condense to produce $\mathcal{C}[\Sigma]$ is non-invertible.

Having introduced the condensation defect, the gauging procedure on $\mathcal{T}(\alpha)[\Sigma]$ described in \cite{Bhardwaj:2022yxj} in order to get the global fusion rule is nothing but stacking $\mathcal{C}[\Sigma]$ on $\mathcal{T}(\alpha)[\Sigma]$, up to a normalization coefficient:
\begin{equation}
\frac{\mathcal{T}(\alpha)[\Sigma]}{\mathbb{Z}_2}\equiv    \frac{1}{\mathcal{Z}(\mathbb{Z}_2;\Sigma)}\mathcal{T}(\alpha)[\Sigma]\otimes \mathcal{C}[\Sigma].
\end{equation}
With this definition, using the fusion of $\mathcal{C}[\Sigma]$ with itself we see that for all the GW operators
\begin{equation}
\label{eq:projection condensation}
    \frac{1}{\mathcal{Z}(\mathbb{Z}_2;\Sigma)}\frac{\mathcal{T}(\alpha)[\Sigma]}{\mathbb{Z}_2}\otimes \mathcal{C}[\Sigma]=\frac{\mathcal{T}(\alpha)[\Sigma]}{\mathbb{Z}_2}.
\end{equation}
The invariance of the GW with $\alpha \neq 0,2\pi$ by stacking $\eta$ is equivalent to
\begin{equation}
\label{eq:projection invariance}
    \mathcal{T}(\alpha)[\Sigma]\otimes \mathcal{C}[\Sigma]=\mathcal{Z}(\mathbb{Z}_2;\Sigma) \mathcal{T}(\alpha) [\Sigma] \ \Rightarrow \ \mathcal{T}(\alpha) [\Sigma]/\mathbb{Z}_2=\mathcal{T}(\alpha) [\Sigma].
\end{equation}
The two equations above can be rephrased by introducing the \emph{projector} $P_{\mathbb{Z}_2}$ which acts on surface operators as
\begin{equation}
 P_{\mathbb{Z}_2}\equiv  \frac{1}{\mathcal{Z}(\mathbb{Z}_2;\Sigma)} \mathcal{C}[\Sigma].
\end{equation}
This is a projector because $P_{\mathbb{Z}_2}^2=P_{\mathbb{Z}_2}$, and we have $P_{\mathbb{Z}_2}\otimes \mathcal{T}(\alpha)[\Sigma]\equiv\mathcal{T}(\alpha)[\Sigma]/\mathbb{Z}_2$. Then \eqref{eq:projection condensation} follows from $P_{\mathbb{Z}_2}^2=P_{\mathbb{Z}_2}$, while \eqref{eq:projection invariance} is just the statement that for the strongly simple objects $\alpha\neq 0,2\pi$, $P_{\mathbb{Z}_2}\otimes \mathcal{T}(\alpha)[\Sigma]=\mathcal{T}(\alpha)[\Sigma]$. On the other hand, the topological operators $\mathcal{T}(0)[\Sigma]/\mathbb{Z}_2$, $\mathcal{T}(2\pi)[\Sigma]/\mathbb{Z}_2$ are further indecomposable objects\footnote{The procedure of adding to the category all the defects obtained by condensations is known in category theory as \emph{idempotent completion}, \emph{Karoubi completion}, or \emph{condensation completion} \cite{Gaiotto:2019xmp, douglas2018fusion}. Objects related among each other by condensation are said to be in the same Schur component, and they have non-trivial morphisms between them.}. This explains why there is not really a mismatch with respect to the 3d case: also in 4d, the defects associated with the short orbits come in different copies obtained by stacking the condensation defect on them. All these copies are connected by 1-morphisms, obtained by putting at the junction lines generating the dual symmetry of the condensed one\footnote{In fusion higher category theory it is known that the simple objects connected by 1-morphisms are only those related among them by condensation \cite{douglas2018fusion}.}. This point will be generalized for $N>2$, but the story will be more involved.

By having understood that to a $\mathbb{Z}_2$ surface operator of the ungauged theory there may correspond different defects of the gauged theory, the necessary modification of the fusion rules, roughly speaking, involves the choices of which of the copies of a given defect appears on the right-hand side. We can determine this by requiring consistency with the fusion with $P_{\mathbb{Z}_2}$: when the left-hand side of the fusion is $P_{\mathbb{Z}_2}$ invariant, also the right-hand side must be invariant. Whenever the local fusion does not have this property, we make it consistent by replacing the right-hand side with $P_{\mathbb{Z}_2}(\mbox{r.h.s.})$. This approach leads to the following modifications (here $\alpha \neq 0,\pi, 2\pi$):
\begin{equation}
    \label{eq:global fusion O(2)}
  \begin{array}{l}
      \mathcal{T}(\alpha)[\Sigma]\otimes \mathcal{T}(2\pi -\alpha)[\Sigma]=2\mathcal{T}(2\pi)[\Sigma]/\mathbb{Z}_2 +\mathcal{T}(2\alpha -2\pi)[\Sigma] \\ \\
       \mathcal{T}(\alpha)[\Sigma]\otimes \mathcal{T}(\alpha)[\Sigma]=2\mathcal{T}(0)[\Sigma]/\mathbb{Z}_2 +\mathcal{T}(2\alpha)[\Sigma]\\ \\
      \mathcal{T}(\pi)[\Sigma]\otimes \mathcal{T}(\pi)[\Sigma]=2\mathcal{T}(0)[\Sigma]/\mathbb{Z}_2+2\mathcal{T}(2\pi)[\Sigma]/\mathbb{Z}_2
  \end{array}  
\end{equation}
in agreement with the fusion rules found in \cite{Bhardwaj:2022yxj}, up to the coefficients in front of the defects on which $\mathbb{Z}_2$ is gauged. This difference boils down to a different normalization for the gauging procedure. Our choice is the one that, when generalized to $N>2$, makes all the fusion coefficients to be positive integer numbers. This makes it possible to relate these coefficients with the total quantum dimensions of the fusion categories of 1-morphisms, made of topological defect lines at the junctions. Indeed in our case, these fusion categories are always categories of modules of finite groups, and they must have integer quantum dimensions equal to the order of the group.
\subsection{$U(1)^{N-1}\rtimes S_N$ Gauge Theory}\label{sec:semiAbelian}
Now we construct the $U(1)^{N-1}\rtimes S_N$ gauge theory we are interested in, by gauging the 0-form symmetry $S_N$ of the Abelian theory. The 3d analog of this theory has been discussed on the lattice in \cite{Nguyen:2021yld}. The toy example in the last subsection has several features of the general case, but there are many other interesting aspects for $N>2$ which make the analysis more complicated. In section \ref{sec:3} we will show the connection between this theory and 4d $SU(N)$ YM theory. For this reason, we present the results in a way to make the comparison with the YM theory suitable.

The local operators are the $S_N$ invariant combinations of those of the Abelian theory, namely all the symmetric polynomials in the $N-1$ variables $\mathcal{F}_{i=1,...,N-1}$. There are $N-1$ independent symmetric polynomials obtained by adding $\mathcal{F}_N=-\mathcal{F}_1-...-\mathcal{F}_{N-1}$ and constructing the $N-1$ symmetric polynomials of degrees $2,3,...,N$ in the $N$ variables $\mathcal{F}_{i=1,...,N}$.

The Wilson lines are the minimal $S_N$ invariant combinations of the Wilson lines $\mathcal{W}(\boldsymbol{n})$ of the Abelian theory $U(1)^{N-1}$. Recalling the action \eqref{eq:action S_n on Wlines}, we construct the Wilson lines of the $U(1)^{N-1}\rtimes S_N$ theory by summing over the orbit of $S_N$
\begin{equation}
    \label{eq:W lines S_N}
    \mathcal{V}(\boldsymbol{n})[\gamma]=\sum _{\sigma \in S_N}\mathcal{W}\left(\mathfrak{S}_{\sigma} ^{\vee} \cdot \boldsymbol{n}\right)[\gamma].
\end{equation}
Now we look at the electric GW operators and their action on the Wilson lines. These are the objects of a 2-category with non-trivial morphisms structure, coming from the dual non-invertible 2-form symmetry $\mbox{Rep}(S_N)$ induced by the gauging of $S_N$. These new topological lines arise as 1-morphisms and play a crucial role in the global fusion. Since the problem is a bit intricate, we start at the local level by putting all the GW on surfaces without non-trivial 1-cycles. We will discuss the 2-category structure and the global fusions in the next subsection. The following discussion applies, \emph{mutatis mutandis}, for the magnetic GW operators as well. The GW operators of the $U(1)^{N-1}\rtimes S_N$ theory are the minimal $S_N$ invariant combination of GW operators of $U(1)^{N-1}$, and their construction is parallel to that for the $S_N$ Wilson lines explained above. We normalize the GW dividing by $|H_{\boldsymbol{\alpha}}|$, where $H_{\boldsymbol{\alpha}}\subset S_N$ is the stabilizer of $\boldsymbol{\alpha}$ 
:
\begin{equation}
\label{eq:GW useful}
\mathcal{T}(\boldsymbol{\alpha})[\Sigma ]=\frac{1}{|H_{\boldsymbol{\alpha}}|}\sum _{\sigma \in S_N}\mathcal{D}\left(\mathfrak{S}_{\sigma}^{\vee}\cdot\boldsymbol{\alpha}\right)[\Sigma].
\end{equation}
These operators are topological and, with same reasons of the $N=2$ case, with this normalization they are always defined as a sum of $\mathcal{D}$ operators without overcounting. 
By construction $\mathcal{T}\left( \mathfrak{S}_{\sigma} ^{\vee} (\boldsymbol{\alpha})\right)=\mathcal{T}(\boldsymbol{\alpha})$, so that the parameter space of the GW operators is
\begin{equation}
    U(1)^{N-1}/ S_N.
\end{equation}
This is a singular manifold since the $S_N$ action on $U(1)^{N-1}$ has fixed points. It is easier to see this in the variables $\beta _i$ introduced above, and we will do it shortly. For the time being we just emphasize that $U(1)^{N-1}/S_N$ coincide with the set of conjugacy classes of $SU(N)$, which labels also the (generically non-topological) GW operators of the $SU(N)$ YM theory \cite{Gukov:2006jk, Gukov:2008sn, Gukov:2014gja}. This is a first clue of a connection between the $U(1)^{N-1}\rtimes S_N$ theory and $SU(N)$ YM theory which we explore in the next section. We will see that it is natural to identify $\mathcal{T}(\boldsymbol{\alpha})$ with the high energy limit of the GW operators of $SU(N)$ YM theory, which becomes topological in the ultraviolet and form a non-invertible symmetry, broken by the RG flow to the center 1-form symmetry $\mathbb{Z}_N^{(1)}$. 

Let us look at the \emph{local fusion} rules . From the definition \eqref{eq:GW useful} we get
    \begin{eqnarray}
     \label{eq:local fusions1}
    \begin{array}{rl}
        \mathcal{T}(\boldsymbol{\alpha})\otimes \mathcal{T}(\boldsymbol{\beta})   & \displaystyle =\frac{1}{|H_{\boldsymbol{\alpha}}||H_{\boldsymbol{\beta}}|} \sum _{\sigma _1,\sigma _2 \in S_N}\mathcal{D}\left(\mathfrak{S}_{\sigma _1}^{\vee} \cdot (\boldsymbol{\alpha}+\mathfrak{S}^{\vee}_{\sigma _1^{-1}\circ \sigma _2}\cdot \boldsymbol{\beta} ) \right) = \\ \\
         & \displaystyle = \frac{1}{|H_{\boldsymbol{\alpha}}||H_{\boldsymbol{\alpha}}|} \sum _{\sigma\in S_N}\sum _{\sigma _1 \in S_N}\mathcal{D}\left(\mathfrak{S}_{\sigma _1}^{\vee} \cdot \left(\boldsymbol{\alpha}+\mathfrak{S}^{\vee}_{\sigma }\cdot \beta \right) \right)= \\ \\
         &\displaystyle = \frac{1}{|H_{\boldsymbol{\alpha}}||H_{\boldsymbol{\beta}}|}\sum _{\sigma \in S_N}|H_{\boldsymbol{\alpha}+\mathfrak{S}^{\vee}_{\sigma }\cdot \boldsymbol{\beta}}|\mathcal{T}\left(\boldsymbol{\alpha}+\mathfrak{S}^{\vee}_{\sigma }\cdot \boldsymbol{\beta} \right)
    \end{array}
\end{eqnarray}
showing that the symmetry generated by the GW operators is non-invertible. This formula is very implicit and does not make it clear the interpretation of the coefficients appearing. Indeed it is important to show that, as in the $N=2$ case, the fusion coefficients are always integer numbers, counting the total quantum dimension of the fusion category of 1-morphisms living at the junctions. We can massage the formula above as follows. Notice that for any $x\in H_{\boldsymbol{\alpha}}, y \in H_{\boldsymbol{\beta}}$ we have $\mathcal{T}\big(\boldsymbol{\alpha}+\mathfrak{S}^{\vee}_{\sigma } \cdot \boldsymbol{\beta}\big)=\mathcal{T}\big(\boldsymbol{\alpha}+\mathfrak{S}^{\vee}_{x\sigma y} \cdot \boldsymbol{\beta}\big)$, and $x\sigma y$ are all the elements of the double coset $H_{\boldsymbol{\alpha}}\sigma H_{\boldsymbol{\beta}}$. Moreover $S_N$ is the disjoint union of all the double cosets, labeled by elements of the double cosets space $H_{\boldsymbol{\alpha}}\backslash S_N /H_{\boldsymbol{\beta}}$. By choosing arbitrarily one element for each double coset the formula above can be rewritten as
\begin{equation}
    \mathcal{T}(\boldsymbol{\alpha})\otimes \mathcal{T}(\boldsymbol{\beta})=\frac{1}{|H_{\boldsymbol{\alpha}}||H_{\boldsymbol{\beta}}|}\sum _{\sigma \in H_{\boldsymbol{\alpha} } \backslash S_N \slash H_{\boldsymbol{\beta}}}|H_{\boldsymbol{\alpha}+\mathfrak{S}^{\vee}_{\sigma }\cdot \boldsymbol{\beta}}| |H_{\boldsymbol{\alpha}}\sigma H_{\boldsymbol{\beta}}| \mathcal{T}\left(\boldsymbol{\alpha}+\mathfrak{S}^{\vee}_{\sigma }\cdot \boldsymbol{\beta} \right) \ .
\end{equation}
The order of the double coset $H_{\boldsymbol{\alpha}}\sigma H_{\boldsymbol{\beta}} $ is \cite{hall2018theory}
\begin{equation}
    |H_{\boldsymbol{\alpha}}\sigma H_{\boldsymbol{\beta}}|=\frac{|H_{\boldsymbol{\alpha}}| |H_{\boldsymbol{\beta}}|}{|H_{\boldsymbol{\alpha}}\cap \sigma H_{\boldsymbol{\beta}} \sigma ^{-1}|}
\end{equation}
from which we find
\begin{equation}
    \label{eq:local fusions}
      \mathcal{T}(\boldsymbol{\alpha})\otimes \mathcal{T}(\boldsymbol{\beta})=\sum _{\sigma \in H_{\boldsymbol{\alpha} } \backslash S_N /H_{\boldsymbol{\beta}}}f_{\alpha \beta}^{\sigma}\;\; \mathcal{T}\left(\boldsymbol{\alpha}+\mathfrak{S}^{\vee}_{\sigma }\cdot \boldsymbol{\beta} \right)
\end{equation}
where
\begin{equation}
    f_{\alpha \beta}^{\sigma} = \frac{|H_{\boldsymbol{\alpha}+\mathfrak{S}^{\vee}_{\sigma }\cdot \boldsymbol{\beta}}|}{|H_{\boldsymbol{\alpha}}\cap \sigma H_{\boldsymbol{\beta}} \sigma ^{-1}|} \in \mathbb{Z}_+ .
\end{equation}
The fusion coefficients $f_{ab}^{\sigma}$ appearing here are integers because $H_{\boldsymbol{\alpha}}\cap \sigma H_{\boldsymbol{\beta}} \sigma ^{-1}$ is a subgroup of $H_{\boldsymbol{\alpha}+\mathfrak{S}^{\vee}_{\sigma }\cdot \boldsymbol{\beta}}$. These numbers are counting the 1-morphisms living at the junctions, up to the endomorphisms. We will shortly see how these numbers are related with the condensation defects that we need to add on right hand side to correct the fusion rules whenever the surface is topologically non-trivial.

Let us look at the Ward identities involving the GW $\mathcal{T}(\boldsymbol{\alpha})[\Sigma]$ and the Wilson lines linking once with $\Sigma$. Consider first a Wilson line $\mathcal{W}(\boldsymbol{n})$ of the $U(1)^{N-1}$ theory, and the action of $\mathcal{T}(\boldsymbol{\alpha})$ on it. By using \eqref{eq:GW on Lines Abelian} we obtain
\begin{equation}
\label{eq: GW semiAbelian on lines}
    \mathcal{T}(\boldsymbol{\alpha})\cdot \mathcal{W}(\boldsymbol{n})=\frac{1}{|H_{\boldsymbol{\alpha}}|}\sum _{\sigma \in S_N}\exp{\left(i\sum _{i,j=1} ^{N-1} \mathfrak{S}_{\sigma} ^{\vee} (\alpha _i) \left(Q^{-1}\right)_{ij}n_j\right)} \mathcal{W}(\boldsymbol{n})=\mathfrak{C} (\boldsymbol{\alpha},\boldsymbol{n})\mathcal{W}(\boldsymbol{n}).
\end{equation}
To prove that the action on the Wilson lines $\mathcal{V}(\boldsymbol{n})$ is diagonal we need to show that $\mathfrak{C} (\boldsymbol{\alpha},\mathfrak{S}_{\sigma}^{\vee} \cdot \boldsymbol{n})=\mathfrak{C} (\boldsymbol{\alpha},\boldsymbol{n})$ for any $\sigma \in S_N$. We recall that from the definition of $Q_{ij}$ we have $\mathfrak{S}_{\sigma} (\mathcal{F}_i)Q_{ij}\mathfrak{S}_{\sigma}(\mathcal{F}_j)=\mathcal{F}_i Q_{ij}\mathcal{F}_j$, implying that $\mathfrak{S}_{\sigma} ^TQ\mathfrak{S}_{\sigma}=Q$. Then $Q^{-1} = \mathfrak{S}_{\sigma}^{-1}Q^{-1}\left(\mathfrak{S}_{\sigma}^{T}\right)^{-1} = \left(\mathfrak{S}_{\sigma}^{\vee}\right)^{T} Q^{-1}\mathfrak{S}^{\vee}_{\sigma}$ which implies $Q^{-1}\left(\mathfrak{S}^{\vee}_{\sigma}\right)^{-1} =\left(\mathfrak{S}_{\sigma}^{\vee}\right)^{T}Q^{-1} $, or $Q^{-1}\left(\mathfrak{S}^{\vee}_{\sigma}\right) =\left(\mathfrak{S}_{\sigma^{-1}}^{\vee}\right)^{T}Q^{-1} $. This gives us the desired invariance
\begin{equation}
\begin{split}
    \mathfrak{C}(\boldsymbol{\alpha}, \mathfrak{S}_{\sigma}^{\vee}\cdot \boldsymbol{n}) &= \frac{1}{|H_{\boldsymbol{\alpha}}|}\sum _{\sigma' \in S_N}\exp{\left(i\boldsymbol{\alpha}^{T}\cdot\left(\mathfrak{S}_{\sigma'} ^{\vee}\right)^{T} Q^{-1}\mathfrak{S}_{\sigma}^{\vee}\cdot \boldsymbol{n}\right)} \\&= \frac{1}{|H_{\boldsymbol{\alpha}}|}\sum _{\sigma ' \in S_N}\exp{\left(i\boldsymbol{\alpha}^{T}\cdot\left(\mathfrak{S}^{\vee}_{ \sigma^{-1} \sigma'}\right)^{T} Q^{-1}\cdot \boldsymbol{n}\right)}  = \mathfrak{C}(\boldsymbol{\alpha}, \boldsymbol{n})
\end{split}
\end{equation}
which proves the following Ward identities
\begin{equation}
\label{eq:actionWilson}
\begin{array}{c}
 \mathcal{T}(\boldsymbol{\alpha})[\Sigma]\cdot \mathcal{V}(\boldsymbol{n})[\gamma]=\mathfrak{C} (\boldsymbol{\alpha},\boldsymbol{n})^{Lk(\Sigma,\gamma)}\mathcal{V}(\boldsymbol{n})[\gamma] \\ \\
  \displaystyle  \mathfrak{C} (\boldsymbol{\alpha},\boldsymbol{n})=\frac{1}{|H_{\boldsymbol{\alpha}}|}\sum _{\sigma \in S_N}\exp{\left(i\sum _{i,j=1} ^{N-1} \mathfrak{S}_{\sigma} ^{\vee} (\alpha _i)\left(Q^{-1}\right) _{ij} n_j\right)}.
\end{array}
\end{equation}
Notice that for $N=2$ we have
$\mathfrak{C} (\alpha,n)=\frac{2}{|H_{\alpha}|}\cos\left(n\frac{\alpha}{2}\right)$, as we obtained before.

The GW operators $\mathcal{T}(\boldsymbol{\alpha})[\Sigma]$ are the generator of a continuous non-invertible symmetry. However an interesting issue is the identification of the sub-category of group-like symmetries. Because the center of $U(1)^{N-1}\rtimes S_N$ is isomorphic to $\mathbb{Z}_N$ we already expect the discrete center symmetry $\mathbb{Z}_N^{(1)}$ to be embedded in the continuous non-invertible symmetry. In the $N=2$ case it was easy to see that $\mathbb{Z}_2$ is the maximal set of invertible unitary generators. We are going to show the same for any $N$, and we provide some interesting property of this center symmetry related with the action on the Wilson lines, to be compared with the non-invertible one. This analysis is also interesting in view of the connection with $SU(N)$ YM theory in the next section, in which only the center symmetry $\mathbb{Z}_N$ remains as an unbroken symmetry along the RG flow.\\
From \eqref{eq:local fusions1} we see that $T(\boldsymbol{\alpha})$ has group-like fusion only if $\boldsymbol{\alpha}$ is a fixed point of the Weyl group. The tricky point here is to properly account for the identifications on the parameters. It is convenient to work in the variables $
    \boldsymbol{\beta} = Q^{-1}\boldsymbol{\alpha}
$ which are separately $2 \pi $ periodic. $S_N$ acts on $\boldsymbol{\alpha}$ with $\mathfrak{S}_{\sigma}^{\vee}$, thus we need to work out the action on  $\boldsymbol{\beta}$. By definition
\begin{equation}
    \beta_{i} = \sum_{j=1}^{N-1}\frac{-1 + N \delta_{ij}}{N}\alpha_{j} = \alpha_{i}-\frac{1}{N}\sum_{j=1}^{N-1}\alpha_{j}
\end{equation}
Since the $\alpha_{i}$ transform in the $\mathfrak{S}^{\vee}$ representation we may write them as $\alpha_{i}= u_{i}-u_{N}$ where $u_{i}$ transform in the $N$-dimensional natural representation. We then have
\begin{equation}
    \beta_{i}  = u_{i}-u_{N}-\frac{1}{N}\sum_{j=1}^{N-1}(u_{j}-u_{N}) =\left(1-\frac{1}{N}\right)u_i  -\frac{1}{N} \sum_{j\neq i} ^{N} u_{j}
\end{equation}
We now introduce an $N$-th variable
\begin{equation}
    \beta_{N} = - \sum_{i=1}^{N-1}\beta_{i} = \left(1-\frac{1}{N}\right)u_{N}-\frac{1}{N}\sum_{j\neq N}u_{j}.
\end{equation}
 Since the $u_{i}$ are permuted by $S_{N}$ it is clear that also the $\beta_{i}$, including $\beta_{N}$, are permuted, i.e. sit in the natural representation. By construction the sum of the $\beta_{i}$ vanishes hence they transform  in the standard $N-1$-dimensional representation. It is now easy to determine the fixed points. Clearly $S_{N}$ contains a subgroup $S_{N-1}$ which permutes the $N-1$ unconstrained $\beta_{i}$'s, those must then be equal at the fixed point: $\beta_{i}=\beta$. The only remaining equation to solve is
\begin{equation}
    \beta  = -\sum_{i=1}^{N-1}\beta  = -(N-1)\beta \ \mbox{mod } 2\pi \ \  \Rightarrow \ \  N \beta = 0  \text{ mod } 2 \pi 
\end{equation}
which is solved by the $N$-th roots of unity
\begin{equation}
    \beta_{*} = \frac{2 \pi k}{N} \quad \quad k = 0,.., N-1.
\end{equation}
This shows that there are $N$ fixed points. We can map them back to the original basis
\begin{equation}
    \alpha_{i} =\sum_{j=1}^{N-1}Q_{ij}\beta_{*} = \sum_{j=1}^{N-1}(1 + \delta_{ij})\beta_{*} = N\beta_{*} = 2 \pi k \quad \quad \forall i = 1,.., N-1.
\end{equation}
We will denote this fixed points by $\boldsymbol{\alpha}_k$, $k=0,...,N-1$. The corresponding fusions are
\begin{equation}
    \mathcal{T}\left(\boldsymbol{\alpha}_k\right) \mathcal{T}\left(\boldsymbol{\alpha}_l\right) = \mathcal{T}\left(\boldsymbol{\alpha}_{l+k}\right)
\end{equation}
proving that these operators form a $\mathbb{Z}_{N}$ subgroup of the non-invertible symmetry. This construction shows that $\mathbb{Z}_{N}$ is the largest possible subcategory with group-like fusions.

Let us now see how this subgroup acts on the lines of the theory. By inserting $\boldsymbol{\alpha}=\boldsymbol{\alpha}_k$ in  (\ref{eq:actionWilson}) we get
\begin{equation}
\begin{split}
    &\mathfrak{C}(\boldsymbol{\alpha}_k,\boldsymbol{n}) = \frac{1}{N!} \sum _{\sigma \in S_N}\exp{\left(i\sum _{i,j=1} ^{N-1} \mathfrak{S}_{\sigma} ^{\vee} (\alpha_i)\left(Q^{-1}\right) _{ij} n_j\right)} = \\
    & = \frac{1}{N!} \sum _{\sigma \in S_N}\exp{\left(i\sum _{i=1} ^{N-1} \mathfrak{S}_{\sigma}(\beta_i)  n_i\right)} =\exp{\left(\frac{2\pi i k}{N}\sum _{i=1} ^{N-1}   n_i\right)}.
    \end{split}
    \label{eq:group_phase}
\end{equation}
This shows that when we restrict to the $\mathbb{Z}_N$ subgroup of the non-invertible symmetry, the action on the Wilson line $\mathcal{V}(\boldsymbol{n})$ becomes group-like with a phase which is an $N$-root of unity with charge 
\begin{equation}
  |\boldsymbol{n}|:=  \sum_{i=1}^{N-1} n_i.
\end{equation} 
\subsection{Higher Condensation and Global Fusion}
\label{sec:global fusion}
When the GW operators are supported on surfaces $\Sigma$ with non-trivial topology we are able to probe the full structure of the 2-category symmetry. An important role is played by the 1-morphisms, which are non-trivial due to the quantum 2-form symmetry arsing by the gauging of $S_N$, implying that there are indecomposable objects with non-trivial endomorphisms. For $N>2$ the quantum symmetry is a discrete non-invertible symmetry $\mbox{Rep}\left(S_N\right)$ and the analysis is more involved with respect to the $O(2)$ gauge theory. The higher condensation defect $\mathcal{C}_{\mbox{Rep}\left(S_N\right)}[\Sigma]$ must be constructed by gauging non-invertible lines on a surface. There is a well established definition of gauging in fusion categories described in \cite{Bhardwaj:2017xup}, and fortunately for any discrete group $G$ the fusion category $\mbox{Rep}(G)$ in 2d can be fully gauged, thus defining the following condensation defect on $\Sigma$:
\begin{equation}
    \mathcal{C}_{\mbox{Rep}\left(S_N\right)}[\Sigma]=\;\;\; \raisebox{-4.5 em}{\includegraphics[width=3cm]{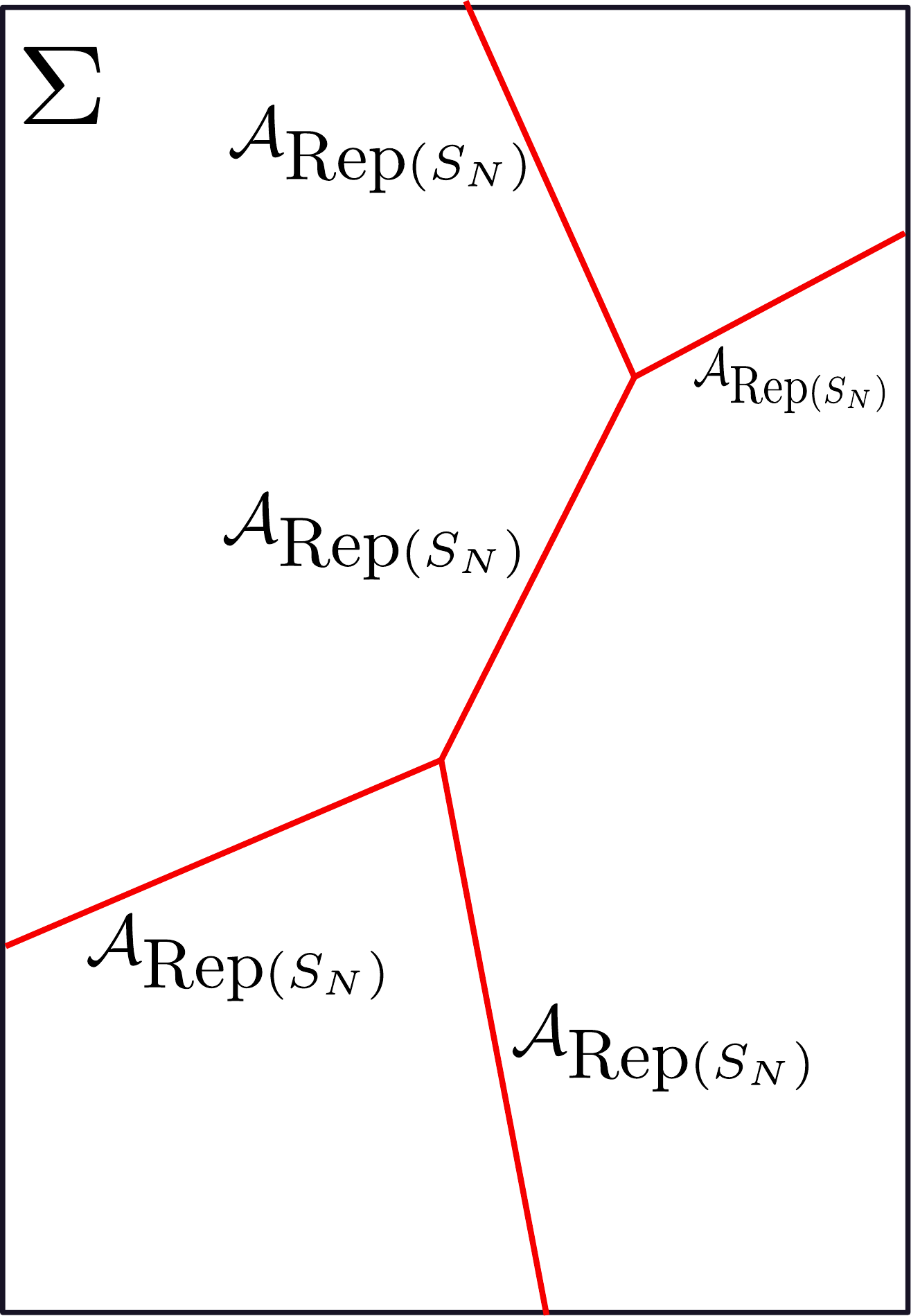}}\;\;\;.
\end{equation}
Here $\mathcal{A}_{\mbox{Rep}\left(S_N\right)}$ is the Frobenius algebra object of $\mbox{Rep}\left(S_N\right)$ corresponding to the \emph{regular} representation\footnote{For Abelian group the regular representation is just the sum of all the irreducible representations, and this generalized gauging procedure coincide with the standard one.}, and by $\mathcal{C}_{\mbox{Rep}\left(S_N\right)}[\Sigma]$ we mean a fine enough mesh of this object on $\Sigma$. On the defect there is a symmetry $S_N$ \cite{Bhardwaj:2017xup}: the fusion category of 1-endomorphisms is the group $S_N$. Notice that the lines generating this symmetry are stacked on the defect and do not exist in the bulk. Below we will give an equivalent description of   $\mathcal{C}_{\mbox{Rep}\left(S_N\right)}[\Sigma]$, which turns out to be useful to compute the fusion with itself, allowing us to find the non-Abelian generalization of the fusions found in \cite{Roumpedakis:2022aik}.

As in the $N=2$ case, we determine the global fusion of $\mathcal{T}(\boldsymbol{\alpha})$ and $\mathcal{T}(\boldsymbol{\beta})$ by requiring consistency with the stacking of Wilson lines which are absorbed by $\mathcal{T}(\boldsymbol{\alpha})$ and $\mathcal{T}(\boldsymbol{\beta})$. For $N=2$ this corresponded to the use of the projector $P_{\mathbb{Z}_2}$, and it was enough because $S_2=\mathbb{Z}_2$ does not have non-trivial proper subgroups: $\alpha \in U(1)/\mathbb{Z}_2$ is either fixed or invariant under $\mathbb{Z}_2$. For $N>2$ there are values $\boldsymbol{\alpha}\in U(1)^{N-1}$ which are stabilized by a non-trivial proper subgroup $H_{\boldsymbol{\alpha}}\subset S_N$. Then the fusion category of 1-endomorphisms $\mathcal{T}(\boldsymbol{\alpha})\rightarrow \mathcal{T}(\boldsymbol{\alpha})$ is isomorphic to $\mbox{Rep}\left(H_{\boldsymbol{\alpha}}\right)$, meaning that the $H_{\boldsymbol{\alpha}}$ Wilson lines are not \emph{absorbed} by $\mathcal{T}(\boldsymbol{\alpha})$ and can live on it as non-trivial lines. On the other hand there are $S_N$ Wilson lines which are not $H_{\boldsymbol{\alpha}}$ Wilson lines, and these are absorbed by $\mathcal{T}(\boldsymbol{\alpha})$. This implies that the local fusion rules require modifications which cannot be seen by simply applying the projector $P_{\mbox{Rep}\left(S_N\right)}$ corresponding to $\mathcal{C}_{\mbox{Rep}(S_N)}[\Sigma]$. Indeed this projector condenses the full symmetry living on the defect:
\begin{equation}
    P_{\mbox{Rep}\left(S_N\right)}\otimes \mathcal{T}\left(\boldsymbol{\alpha}\right)[\Sigma]=\mathcal{T}\left(\boldsymbol{\alpha}\right)[\Sigma]/\mbox{Rep}\left(H_{\boldsymbol{\alpha}}\right).
\end{equation}
When $\mathcal{T}(\boldsymbol{\alpha})$ is a strongly simple object, namely $H_{\boldsymbol{\alpha}}=1$, by using $P_{\mbox{Rep}\left(S_N\right)}$ we can determine the correct fusion rules. On the other hand if $H_{\boldsymbol{\alpha}}$ is a non-trivial proper subgroup, using only $P_{\mbox{Rep}\left(S_N\right)}$ we would miss the global fusion rules with $\mathcal{T}\left(\boldsymbol{\alpha}\right)[\Sigma]$ appearing on the left hand side. We then need to construct the projector containing the maximal set of lines absorbed by $\mathcal{T}(\boldsymbol{\alpha})$. Before clarifying what does this mean and giving a general construction, we need to introduce the promised alternative definition of $\mathcal{C}_{\mbox{Rep}(S_N)}[\Sigma]$.

The idea is that since the $\mbox{Rep}(S_N)$ symmetry is obtained by the gauging of $S_N$, condensing it on $\Sigma$ is the same as doing a step back before gauging $S_N$, removing $\Sigma$ from the space-time manifold $\mathcal{M}$ and then gauging $S_N$ in $\mathcal{M}-\Sigma$. We do so imposing Dirichlet boundary conditions $a|_{\Sigma}=0$ on the surface for the $S_N$ gauge field $a$. This construction produces the $U(1)^{N-1}\rtimes S_N$ theory with the insertion of a condensation defect $\mathcal{C}_{\mbox{Rep}(S_N)}[\Sigma]$. Notice that this picture is consistent with the presence of a dual $S_N$ symmetry living on $\mathcal{C}_{\mbox{Rep}(S_N)}[\Sigma]$: a co-dimension one defect of the 0-form global symmetry $S_N$ in the $U(1)^{N-1}$ theory can intersect $\Sigma$ on a line wrapping a cycle, then this defect is made transparent outside $\Sigma$ by the gauging of $S_N$ in $\mathcal{M}-\Sigma$, while the line on $\Sigma$ remains as the generator of a 0-form symmetry on the condensation defect.

\begin{figure}[t!]
 \centering
   \raisebox{-0 em}{\includegraphics[width=15cm]{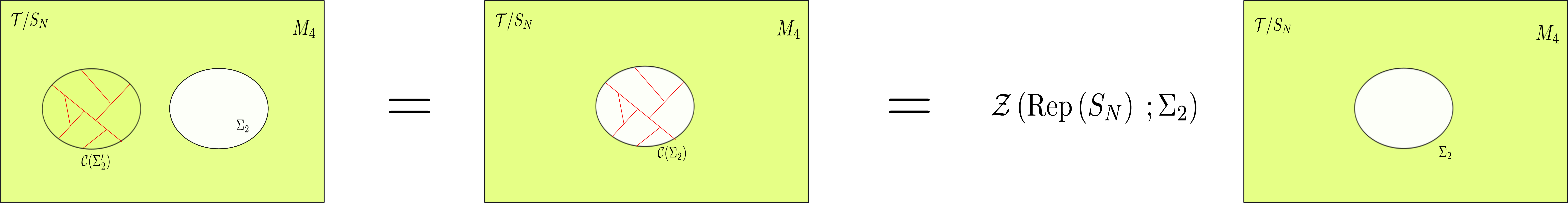}}
    \caption{A pictorial representation of the fusion of the condensation defects. The green region is the one with a gauged $S_N$ symmetry while the white one is the one in which such a symmetry is still global. The red lines represent a fine enough mesh of the algebra object representing the gauging of Rep($S_N$) in the $2$-dimensional surface $\Sigma_2$ or $\Sigma'_2$. }
  \label{fig:condensation_fusion}
\end{figure}

This way of presenting $\mathcal{C}_{\mbox{Rep}(S_N)}[\Sigma]$ may seem abstract, but it does not rely on the concept of gauging a Frobenius algebra object, and turns out to be useful to determine the fusion  $\mathcal{C}_{\mbox{Rep}(S_N)}[\Sigma]\otimes \mathcal{C}_{\mbox{Rep}(S_N)}[\Sigma]$. For convenience we denote the defect constructed in this way by $\widetilde{\mathcal{C}}$, even if $\widetilde{\mathcal{C}}=\mathcal{C}$, to distinguish when we are thinking about the condensation defect in the standard or in the latter presentation. To compute $\mathcal{C}_{\mbox{Rep}(S_N)}[\Sigma]\otimes \mathcal{C}_{\mbox{Rep}(S_N)}[\Sigma]$ the trick is to think one of the two supported on $\Sigma$ and defined in the presentation $\widetilde{\mathcal{C}}$, while the other defined in the standard way $\mathcal{C}$ with the condensation of the algebra object $\mathcal{A}_{\mbox{Rep}(S_N)}$ on a surface $\Sigma '=\Sigma +\delta \Sigma$, which lies inside the mesh of $S_N$ defects in $\mathcal{M}-\Sigma$. When we send the displacement $\delta \Sigma$ to zero the mesh of $\mathcal{A}_{\mbox{Rep}(S_N)}$ defining $\mathcal{C}$ enters into the "hole" $\Sigma$ defining $\widetilde{\mathcal{C}}$ (see figure \ref{fig:condensation_fusion}). The result is again $\widetilde{\mathcal{C}}_{\mbox{Rep}(S_N)}[\Sigma]$ but with the hole $\Sigma$ filled with a mesh of algebra objects implementing the higher gauging of $\mbox{Rep}(S_N)$. Because of the Dirichlet boundary conditions this condensation does not speak with the $S_N$ gauge field in the bulk, and it simply computes the partition function of the 2d pure $\mbox{Rep}(S_N)$ gauge theory on $\Sigma$, denoted by $\mathcal{Z}\left( \mbox{Rep}\left(S_N\right) ; \Sigma \right) $. Since $\widetilde{\mathcal{C}}=\mathcal{C}$ we get
\begin{equation}
    \mathcal{C}_{\mbox{Rep}(S_N)}[\Sigma]\otimes \mathcal{C}_{\mbox{Rep}(S_N)}[\Sigma]=\mathcal{Z}\left( \mbox{Rep}\left(S_N\right) ; \Sigma \right) \mathcal{C}_{\mbox{Rep}(S_N)}[\Sigma]
\end{equation}
which can be thought of as a non-Abelian generalization of the results in \cite{Roumpedakis:2022aik}. The pure $\mbox{Rep}(S_N)$ gauge theory is a theory with non-abelian 0-form symmetry $S_N$, which can be described explicitly in terms of commutative Frobenius algebras. In appendix \ref{sec:2dTQFT} we provide some detail on this construction.

We check the correctness of this abstract procedure by repeating it in an Abelian case where $S_N$ is replaced by $\mathbb{Z}_N$. This has the advantage that the dual symmetry is invertible and its higher gauging on $\Sigma ' \subset \mathcal{M}-\Sigma$ can be done by simply coupling the defect to a background gauge field $b \in H^1\left(\Sigma ' , \mathbb{Z}_N\right)$ and summing over it. The coupling of $b$ to the $\mathbb{Z}_N$ gauge field $a\in H^1(\mathcal{M}-\Sigma, \mathbb{Z}_N)$ is the standard one:
\begin{equation}
\label{eq:coupling}
  \exp{\left(\int _{\Sigma '} a\cup b\right)}\mathcal{Z}(\mathbb{Z}_N; \Sigma ').
\end{equation}
By summing over $b$ one gets the condensation defect in the standard presentation $\mathcal{C}$ of \cite{Roumpedakis:2022aik}. On the other hand our alternative definition $\widetilde{\mathcal{C}}$ is formally the same in the Abelian and in the non-Abelian case, since it does not use the notion of gauging non-invertible symmetries. The insertion of $\mathcal{C}_{\mathbb{Z}_N}[\Sigma]\otimes \mathcal{C}_{\mathbb{Z}_N}[\Sigma] $ in a correlation function can be replaced by \eqref{eq:coupling} in the same correlation function, computed in the theory where $\mathbb{Z}_N$ is gauged in $\mathcal{M}-\Sigma$ with Dirichlet boundary condition $a|_{\Sigma}=0$, and then take the limit $\Sigma '\rightarrow \Sigma$. In this limit the exponential factor disappear because of the Dirichlet boundary condition. We remain with the partition function of the 2d $\mathbb{Z}_N$ gauge theory on $\Sigma$ multiplying the correlation function computed in the theory with dynamical gauge field $a\in H^1(\mathcal{M}-\Sigma,\mathbb{Z}_N)$. This means
\begin{equation}
    \mathcal{C}_{\mathbb{Z}_N}[\Sigma]\otimes \mathcal{C}_{\mathbb{Z}_N}[\Sigma]=\mathcal{Z}\left(\mathbb{Z}_N; \Sigma \right) \mathcal{C}_{\mathbb{Z}_N}[\Sigma]  
\end{equation}
which is the same fusion of \cite{Roumpedakis:2022aik}. 

The non-Abelian condensation defects we defined allow to construct the projector $P_{\mbox{Rep}(S_N)}$ satisfying $P_{\mbox{Rep}(S_N)}^2=P_{\mbox{Rep}(S_N)}$. By using it we obtain the global fusions of the strongly simple GW operators, namely those with trivial stabilizers $H_{\boldsymbol{\alpha}}=H_{\boldsymbol{\beta}}=1$
\begin{equation}
\label{eq:global fusion with trivial stabilizer on the lhs}
\begin{array}{rr}
    \mathcal{T}(\boldsymbol{\alpha})[\Sigma]\otimes  \mathcal{T}(\boldsymbol{\beta})[\Sigma] &  \displaystyle =\sum _{\sigma \in S_N} |H_{\boldsymbol{\alpha} +\mathfrak{S}_{\sigma} ^{\vee} \cdot \boldsymbol{\beta}}|P_{\mbox{Rep}(S_N)}\otimes  \mathcal{T}\left(\boldsymbol{\alpha} +\mathfrak{S}_{\sigma} ^{\vee} \cdot \boldsymbol{\beta}\right)[\Sigma]=\\
    \\
    &\displaystyle =\sum _{\sigma \in S_N} |H_{\boldsymbol{\alpha} +\mathfrak{S}_{\sigma} ^{\vee} \cdot \boldsymbol{\beta}}| \frac{\mathcal{T}\left(\boldsymbol{\alpha} +\mathfrak{S}_{\sigma} ^{\vee} \cdot \boldsymbol{\beta}\right)[\Sigma]}{\mbox{Rep}\left(H_{\boldsymbol{\alpha} +\mathfrak{S}_{\sigma} ^{\vee} \cdot \boldsymbol{\beta}}\right)}
\end{array}
\end{equation}
where we used that the projector on the right hand side implements the gauging of the full symmetry $\mbox{Rep}\left(H_{\boldsymbol{\alpha} +\mathfrak{S}_{\sigma} ^{\vee} \cdot \boldsymbol{\beta}}\right)$ living on the GW $\mathcal{T}\left(\boldsymbol{\alpha} +\mathfrak{S}_{\sigma} ^{\vee} \cdot \boldsymbol{\beta}\right)$.

As advertised before, when $H_{\boldsymbol{\alpha}}$ is a non-trivial proper subgroup of $S_N$ we need the maximal projector absorbed by $\mathcal{T}(\boldsymbol{\alpha})$. A priory it is not obvious at all how to define this projector. If $H_{\boldsymbol{\alpha}}$ is a normal subgroup, then $H_{\boldsymbol{\alpha}}^{\perp}=S_N/H_{\boldsymbol{\alpha}}$ is a group, and intuitively we need a projector $P_{\mbox{Rep}(H_{\boldsymbol{\alpha}}^{\perp})}$ obtained from the condensation of $\mbox{Rep}(H_{\boldsymbol{\alpha}}^{\perp})$ Wilson lines. However it is not obvious that this an allowed gauging in the category $\mbox{Rep}(S_N)$ of bulk lines, and more seriously we would not know how to proceed when the stabilizer is not a normal subgroup\footnote{For $N\geq 5$ there is only one non-trivial proper normal subgroup of $S_N$, namely the alternating group $A_N$.}. Our definition of the relevant condensation defect absorbed by $\mathcal{T}(\boldsymbol{\alpha})$ is as follows. We start from the maximal condensation defect $\mathcal{C}_{\mbox{Rep}(S_N)}[\Sigma]$ and we recall that there is a quantum symmetry $S_N$ living on it, which is very explicit in our presentation $\widetilde{\mathcal{C}}$ of this defect. Then for any subgroup $H_{\boldsymbol{\alpha}}\subset S_N$ we can gauge this smaller symmetry on the defect, which corresponds to remove the $\mbox{Rep}(H_{\boldsymbol{\alpha}})$ Wilson line from the condensate, and generate a \emph{new} higher-condensation defect which, with an abuse of notation, we denote with $\mathcal{C}_{\mbox{Rep}(H_{\alpha}^{\perp})}[\Sigma]$. Notice that this construction matches nicely with the known fact that the Frobenius algebra objects of $\mbox{Rep}(G)$ are in one-to-one correspondence with the subgroups of $G$ \cite{Bhardwaj:2017xup}.

From this defect we can construct the projector $P_{\mbox{Rep}(H_{\boldsymbol{\alpha}}^{\perp})}$ for any $\boldsymbol{\alpha} \in U(1)^{N-1}/S_N $, and this is the maximal projector absorbed by $\mathcal{T}(\boldsymbol{\alpha})$. When we fuse two of these higher condensation defects for $\boldsymbol{\alpha}$ and $\boldsymbol{\beta}$, we are essentially removing from the condensate $\mathcal{C}_{\mbox{Rep}(S_N)}[\Sigma]$ all the lines which are lines of both $H_{\boldsymbol{\alpha}}$ and $H_{\boldsymbol{\beta}}$, while keeping all the others. This leads to the following algebra of projectors
\begin{equation}
\label{eq:projectors algebra}
    P_{\mbox{Rep}(H_{\boldsymbol{\alpha}}^{\perp})}\otimes P_{\mbox{Rep}(H_{\boldsymbol{\beta}}^{\perp})}=P_{\mbox{Rep}\left(\left(H_{\boldsymbol{\alpha}}\cap H_{\boldsymbol{\beta}}\right)^{\perp}\right)}
\end{equation}
We can use this knowledge to compute the most general global fusion rules, by starting from the local one \eqref{eq:local fusions} and apply the projectors $P_{\mbox{Rep}(H_{\boldsymbol{\alpha}}^{\perp})}$ and $P_{\mbox{Rep}(H_{\boldsymbol{\beta}}^{\perp})}$ to both sides of the equation, which are absorbed by the left-hand side:
\begin{equation}
\label{eq:general global fusion}
\mathcal{T}(\boldsymbol{\alpha})[\Sigma]\otimes  \mathcal{T}(\boldsymbol{\beta})[\Sigma]   =\sum _{\sigma \in H_{\boldsymbol{\alpha}}\backslash S_N /H_{\boldsymbol{\beta}}}f_{\alpha \beta}^{\sigma} \; P_{\mbox{Rep}\left(\left(H_{\boldsymbol{\alpha}}\cap H_{\boldsymbol{\beta}}\right)^{\perp}\right)}\otimes  \mathcal{T}\left(\boldsymbol{\alpha} +\mathfrak{S}_{\sigma} ^{\vee} \cdot \boldsymbol{\beta}\right)[\Sigma] \;\;\; .
\end{equation}
It is a trivial exercise to check that this formula agrees with the global fusion of the $O(2)$ gauge theory. The general fusion rule above explains the meaning of the integer fusion coefficients coefficients $f_{ab}^{\sigma}$. These numbers are greater than one whenever they multiply an operator dressed with some condensation defect, and the number coincide with the quantum dimension of the algebra object condensed on the defect. This fact has a simple interpretation. The condensation produces a dual symmetry on the defect, and a junction among the two fused defects and any one of those appearing on the right can be constructed using any of the lines generating this dual symmetry, whose total quantum dimension is equal to that of the condensed algebra object.

We can look back at the $N=2$ case and check that this discussion applies. A richer example is the case $N=3$, and it is worth to discuss it here. Given $\boldsymbol{\alpha}=(\alpha _1,\alpha _2)\in U(1)^2/S_3$ there are four possible stabilizers:
\begin{itemize}
    \item $\alpha _1=\alpha _2=2\pi k$, $k=0,1,2$ is fixed by the full group $H_{\boldsymbol{\alpha}}=S_N$.
    \item For $\alpha _1=2\pi k_1$, $\alpha _2=2\pi k_2 $, $k_1,k_2=0,1,2$, $k_1\neq k_2$ the stabilizer is $H_{\boldsymbol{\alpha}}=\mathbb{Z}_3$.
    \item For $\alpha _1=\alpha _2=:\alpha$, but $\alpha /2\pi \notin \mathbb{Z}$ the stabilizer is $H_{\boldsymbol{\alpha}}=\mathbb{Z}_2$
    \item In all the other cases the stabilizer is trivial.
\end{itemize}
When $\boldsymbol{\alpha}=(\alpha,\alpha)$, $\boldsymbol{\beta}=(\beta,\beta)$ are both stabilized by $\mathbb{Z}_2$, by using \eqref{eq:local fusions} the local fusion is\footnote{The fact that only two terms appear in the right hand side follows the Cauchy-Frobenius lemma
\begin{equation*}
|H_1\backslash G/H_2|=\frac{1}{|H_1||H_2|}\sum _{h_1\in H_1, h_2\in H_2} |G_{h_1,h_2}|    
\end{equation*}
where $G_{h_1,h_2}=\left\{g\in G \; | \; h_1gh_2=g\right\}$. Indeed $\mathbb{Z}_2=\left\{1,s\right\}\subset S_3$, where $s=(213)$, and it is easy to see that $G_{11}=S_3$, $G_{1x}=G_{x1}=\emptyset$, $G_{xx}=\mathbb{Z}_2$.}
\begin{equation}
 \mathcal{T}\left(\alpha ,\alpha \right)\otimes    \mathcal{T}\left(\beta ,\beta \right)=\mathcal{T}\left(\alpha ,\alpha-\beta \right)+\frac{|H_{\boldsymbol{\alpha}+\boldsymbol{\beta}}|}{2}\mathcal{T}\left(\alpha +\beta  ,\alpha +\beta \right)
\end{equation}
which is modified, at the global level, by gauging $\mathbb{Z}_3$. Since we are assuming $\alpha ,\beta \notin 2\pi \mathbb{Z}$ the first term cannot be stabilized by $\mathbb{Z}_3$. Then the only non-trivial modification is when $\beta=2\pi k-\alpha$, $\alpha /2\pi \notin \mathbb{Z}$, in which case the last term is central, and we get
\begin{equation}
     \mathcal{T}\left(\alpha ,\alpha \right)\otimes    \mathcal{T}\left(2\pi k-\alpha ,2\pi k-\alpha \right)=\mathcal{T}\left(\alpha ,2\alpha-2\pi k \right)+3\frac{\mathcal{T}\left(2\pi k  ,2\pi k \right)}{\mathbb{Z}_3} \ .
\end{equation}
Notice that, even if the last GW in the local fusion is a generator of the center which stabilizes the full $S_3$, the condensation defect dressing it in the global fusion is the one associated with $\mathbb{Z}_3$. There is a quantum $\mathbb{Z}_3$ symmetry on this defect, and the coefficient 3 is counting precisely its total quantum dimension.

Now we fuse two GW whose parameters $\boldsymbol{\alpha}_{a_1a_2}=(2\pi a_1,2\pi a_2), \boldsymbol{\beta}_{b_1b_2}=(2\pi b_1,2\pi b_2)$ have both stabilizer $\mathbb{Z}_3$. The local fusion is
\begin{equation}
    \mathcal{T}(\boldsymbol{\alpha}_{a_1a_2})\otimes \mathcal{T}(\boldsymbol{\beta}_{b_1b_2})=\frac{|H_{\boldsymbol{\alpha}_{a_1a_2}+\boldsymbol{\beta}_{b_1b_2}}|}{3} \mathcal{T}(\boldsymbol{\alpha}_{a_1a_2}+\boldsymbol{\beta}_{b_1b_2})+\frac{|H_{\boldsymbol{\alpha}_{a_1a_2}+\boldsymbol{\beta}_{b_2b_1}}|}{3} \mathcal{T}(\boldsymbol{\alpha}_{a_1a_2}+\boldsymbol{\beta}_{b_2b_1})
\end{equation}
which should be modified by applying $P_{\mathbb{Z}_2}$. Notice that it is impossible that both terms on the right hand side are stabilized by $\mathbb{Z}_2$, otherwise $a_1=a_2, b_1=b_2$. When the second term is stabilized by $\mathbb{Z}_2$ we get the global fusion rule
\begin{equation}
\begin{array}{r}
   \displaystyle \mathcal{T}(2\pi a_1,2\pi a_2)\otimes \mathcal{T}\left(2\pi b_1,2\pi (a_2+b_1-a_1)\right)= \mathcal{T}(2\pi (a_1+b_1),2\pi (2a_2+b_1-a_1))+  \\ \\  \displaystyle + 2\frac{\mathcal{T}(2\pi (a_2+b_1),2\pi (a_2+b_1))}{\mathbb{Z}_2}
\end{array}
\end{equation}
The coefficient 2 in the last term has the same interpretation of the 3 in previous case, as the number of possible junctions.

Because $S_3=\mathbb{Z}_3\rtimes \mathbb{Z}_2$, this example can be analyzed also with the technique of gauging sequentially $\mathbb{Z}_3$ and then $\mathbb{Z}_2$ as in  \cite{Bhardwaj:2022yxj}, and one can check that we reproduce the same global fusions. On the other hand our method is more general since it does not assume that the group to be gauged is a semidirect product of Abelian factor, which is not true for $S_N$, $N\geq 5$. Nevertheless the computation is incredibly harder for $N>3$, even if it is algorithmic.

We conclude this subsection with a general remark. The method we described to derive the global fusion rules in the $U(1)^{N-1}\rtimes S_N$ appears to be general in higher category symmetries. The difference between local and global fusion arises in this context because also indecomposable objects can have a non-trivial category of 1-endomorphisms, and one needs to require consistency of the fusions with the condensation of these symmetries generated by 1-endomorphisms. This takes the form of various projections obtained by fusing with the higher condensation defects introduced in \cite{Roumpedakis:2022aik}. As we have discussed, the determination of the full set of higher condensation defects of a given theory might be non-trivial. Nevertheless we propose that, at least for non-invertible symmetries induced by gauging, the only modification of the local fusions required when the defects have non-trivial topology are those coming from these consistency conditions. As a consequence, finding all the higher condensation defects of a theory allows to fully determine the global fusions. This proposal is motivated by the observation that the only difference arising when the defect is topologically non-trivial can be in the presence of lower dimensional defects wrapping cycles. By definition, the higher condensation defects are precisely those which are made by lower dimensional objects\footnote{It is worth noting that this procedure is reminiscent of the idempotent completion in higher categories introduced in \cite{Gaiotto:2019xmp}. It would be very interesting to draw a precise connection with the recent known mathematical results}. Notice, for example, that the way in which the authors of \cite{Choi:2021kmx} determined the fusion rules of the \emph{duality defects}, after being aware of condensation defects, can be interpret as our method. 
\section{The Ultraviolet Limit of 4d Yang-Mills Theory}
\label{sec:3}
In this section we connect the $U(1)^{N-1}\rtimes S_N$ gauge theories to $SU(N)$ YM theories showing that \emph{some} properties of the UV limit of the latter are nicely captured by the former. We will argue that a convenient way to analyze this relation boils down to choosing a particular gauge fixing, originally introduced in \cite{tHooft:1981bkw}, in which the connection with the semi-Abelian theory is manifest. We will show that all gauge invariant operators of $SU(N)$ YM theory are matched by operators in the semi-Abelian theory. The relation we find implies that the global symmetries of the high energy YM theory are much larger than those of the full theory as they include much more topological operators which generate a non-invertible symmetry\footnote{The same conclusion was argued for $SU(2)$ YM theory in \cite{Cordova:2022rer} by directly doing the $g \rightarrow 0$ limit.}.

Naively one might say that the UV limit of $SU(N)$ YM theory is sharply different from the $U(1)^{N-1}\rtimes S_N$, since the latter is locally a theory of $N-1$ photons, while the former seems to be a theory of $N^2-1$ free gluons. However in a non-Abelian theory there are much more gauge transformations than in a collection of Abelian ones, as for instance gluons can be rotated into each other, thus a UV description in terms $N^2-1$ photons is misleading as it does not account for all the redundancies. In order to introduce the general idea, it is useful to look at a toy example. Consider the matrix model of $N\times N$ hermitian matrices. Here it is clear that, by diagonalizing the matrices, we can reduce the initial $N^{2}$ degrees of freedom to only $N$ at the price of introducing a potential among them related to the Vandermonde determinant\footnote{Note that in this diagonal gauge the Weyl group which permutes the eigenvalues is still a gauge symmetry.} (for a review see e.g. \cite{Marino:2012zq}). This determinant is crucial in order to match all the calculable quantities of the original theory\footnote{For instance the free energy computed directly from the matrix model is proportional to $g^{N^2}$, signal of the fact that the theory contains $N^2$ degrees of freedom. The theory described by the eigenvalues gives the same result only if the Vandermonde is taken into account.}. However the gauge invariant content of the theory is entirely captured by the $N$ degrees of freedom and, in the free limit, the measure induced potential are turned off and become irrelevant to study \emph{kinematical} properties of the original theory.

Inspired by this example we can argue that the UV limit of $SU(N)$ YM theory is related to the semi-Abelian gauge theory $U(1)^{N-1} \rtimes S_N$. In particular, even if the dynamics of the YM theory at arbitrary small coupling is \emph{not} captured by the semi-Abelian theory, the equations of motion of the latter, together with the symmetry structure, carry over to the UV limit of the YM theory. In the next subsection we make this argument more precise. Then the other subsections are devoted to show the matching of all the gauge invariant operators. Finally we will discuss how the possible global structures of the YM theory are captured by the semi-Abelian theory.

\subsection{Yang-Mills theory}
\label{sec: YM}
Consider the 4d $SU(N)$ YM theory
\begin{equation}
    Z = \int DA e^{-\frac{1}{2g^2}\int d^4x Tr F \wedge *F}
\end{equation}
where $F = dA + A \wedge A$ is an hermitian and traceless matrix transforming covariantly under $SU(N)$ gauge transformations $
    F \rightarrow \Omega^{-1}F\Omega
$. We use the letters  $i,j,...$ for the generators $h_i$ in the Cartan subalgebra, while $a,b,...$ for the off-diagonal ones $T_a$.
We use the non-Abelian gauge redundancy to choose a gauge in which $F$ is diagonal $F=F_ih_i$ \footnote{The idea to Abelianize a non-Abelian gauge theory using particular gauge fixing was introduced in \cite{tHooft:1981bkw}. In particular this method was made rigorous and it was used in \cite{Blau:1995rs} in order to solve $G$ YM theories in $2$d where these theories are quasi-topological and solvable.}. In this gauge the action of the theory becomes 
\begin{equation}
    S = \frac{1}{2g^2} \int d^4x \sum_{i,j=1} ^{N-1} K_{ij}F_i\wedge * F_j
\end{equation}
and all the complicated dynamics is then captured by the induced gauge fixing determinant.
\\
The $(N-1)\times (N-1)$ matrix $K_{ij}$ is the Killing form restricted to the Cartan subalgebra.  It is useful to choose the Chevalley basis in which
\begin{equation}
    h_i=2\frac{\alpha _i ^I  H^I}{|\alpha _i|^2}  \ \ \ \ K(H^I,H^J)=\delta ^{I,J}.
\end{equation}
so that (for simply-laced Lie algebras) the Killing form restricted to the Cartan subalgebra is the Cartan matrix
\begin{equation}
   K_{ij}=\frac{2}{|\alpha _i|^2}\frac{2\alpha _i ^I \alpha _j^JK(H^I,H^J)}{|\alpha _j|^2}=2\frac{\alpha _i \cdot \alpha _j}{|\alpha _j|^2}=A_{ij}.
\end{equation}
The residual gauge freedom is now described by the semi-Abelian gauge group $U(1)^{N-1} \rtimes S_N$, where $U(1)^{N-1}$ is the maximal torus of $SU(N)$ and $S_N$ is the Weyl group. Its gauging reflects the freedom of defining the $N$ eigenvalues $F_i$ in different orders. 

As opposed to the simpler case of the matrix model, this gauge fixing condition is now more complicated. In what follows we sketch how this procedure should be done, even if in order to analyze the kinematical properties of the UV theory (such as the symmetries) all the technicalities turn out not to be crucial. 
\\ In the YM path integral we integrate over the connections, not over the field strengths which are the objects transforming covariantly. However we can still do similar considerations. The gauge fixing condition which we want to impose is
\begin{equation}
    F^a = 0.
    \label{Abeliangaugefixing}
\end{equation}
Usually in the Faddeev-Popov procedure we do not resolve the $\delta$-function corresponding to the gauge fixing, but instead we rewrite it as a gauge fixing term in the action. In this case however it is convenient to resolve the $\delta$-function, so that the constraint is imposed directly in the action. This is because we do not want to preserve the full gauge covariant form of the action, but only the $U(1)^{N-1}$ one.
Note that the connections $A^a$ are not necessarily zero, only their field strengths are. There is an induced Faddeev-Popov determinant so that the gauge-fixed path integral looks like
\begin{equation}
    Z = \int DA^i \;DA^a e^{-\frac{1}{2}\sum_{i,j} F_iF_j K_{ij}} \Delta(A^i,A^a).
\end{equation}
In writing this we used the normalization in which the field strength is $F = dA + g A\wedge A$ so that in the $g=0$ limit we just get the Abelian kinetic term for the $A^i$ connections\footnote{Note that in this case also in $\Delta(A^i,A^a)$ there is a dependence on $g$. }. When we write $\Delta(A^i,A^a) = e^{V(A^i,A^a)}$ and we integrate over $A^a$, we induce complicated non-local interactions among the Cartan gauge connections $A^i$. These interactions play the same role as the Vendermonde determinant, and in particular it will be crucial in order to match all the complicated dynamics of the non Abelian theory, precisely as in the matrix model. However these interactions are weighted by the gauge coupling $g$, and in the high energy limit are turned off. Therefore, for what concerns the analysis of gauge invariant operators and symmetries we can safely drop the non-local interactions at high energy and study the remaining theory, which is precisely the $U(1)^{N-1}\rtimes S_N$ gauge theory.

We want now to discuss an additional issue which concerns the global properties of the non Abelian theory. Indeed in the $SU(N)$ theory we have different instanton sectors labeled by the third homotopy group of the gauge group. When we fix the gauge we loose this information since the residual gauge symmetry has no non-trivial topological sectors. This means that this gauge fixing works only locally and it must be modified if we want to account the global properties of the theory \cite{Blau:1994rk}. However, in the $g=0$ limit also in the $SU(N)$ theory all the non-trivial instanton sectors decouple and the lack of non-trivial topological sectors is no longer an issue.

The argument above suggests to look for a mapping between the gauge invariant operator of the $SU(N)$ YM theory and those of the semi-Abelian theory, in the following three subsections we show the precise correspondence. We want to stress that the matching of the gauge invariant operators is independent of the gauge fixing procedure since it comes just from the freedom of applying gauge rotations on gauge invariant quantities. Instead, the power of these considerations is that, once we understand the map of operators, we will be able to extract some information about the UV limit of YM theory knowing the properties of the semi-Abelian one already discussed in the previous section. In the YM theory the observables are of three kinds:
\begin{itemize}
    \item \textbf{Local operators.} In pure YM theory they are all constructed out of the field strength. The most obvious one are the Lagrangian itself $Tr(F\wedge *F)$ and the $\theta -$term $Tr(F\wedge F)$, but there are others like $Tr((F\wedge * F)\wedge *(F\wedge *F))$ and so on.
    \item \textbf{Line operators.} These are the simplest kind of extended operators, supported on lines. In YM theory they are the Wilson operators
    \begin{equation}
        W_{\mathcal{R}}[\gamma]=Tr _{\mathcal{R}} P \exp{\left(i \oint _{\gamma} A \right)}
    \end{equation}
    labeled by an irreducible representation of the gauge group, as well as the 't Hooft lines, defined as disorder operators \cite{tHooft:1977nqb} . These are also labeled by representations \cite{Kapustin:2005py}, but for gauge group $SU(N)$ only those with $N-$ality zero are \emph{genuine line operators}, while the others require topological surfaces attached to them \cite{Aharony:2013hda}.
    \item \textbf{Surface operators.} These operators are supported on surfaces, which in 4d can link with lines $Lk(\Sigma, \gamma)\in \mathbb{Z}$. For this reason there is a crucial interplay between line and surface operators. When a surface operator is topological it is a generator of a 1-form symmetry, possibly non-invertible, and the charged objects are line operators. In 4d gauge theories the surface operators are known as GW operators \cite{Gukov:2006jk,Gukov:2008sn,Gukov:2014gja}. They are labeled by the conjugacy classes of $G$ parametrized by $U(1)^r/W_G$, but only those corresponding to the center $Z(G)\subset G$ are topological in the full theory, generating the center symmetry. 
\end{itemize}

In the next three subsections we will discuss the matching of the three types of operator between the $U(1)^{N-1}\rtimes S_N$ gauge theory and the UV effective description of YM theory. In the  discussion on local operator (subsection \ref{sec:local operators}) we also clarify the relation between the various Lagrangians, in the various basis.
\subsection{Local Operators}
\label{sec:local operators}
As explained above, in the $g_{YM}\rightarrow 0$ limit we can reduce to the action
\begin{equation}
\label{eq:action UV YM}
    S=\frac{1}{2}\int d^4x \; K_{ij}F_i\wedge * F_j
\end{equation}
where $K_{ij}=K(h_i,h_j)$ is the block of the Killing form relative to the Cartan subalgebra. This is an Abelian gauge theory with gauge group $U(1)^{N-1}$. As pointed out in section (\ref{sec:Abeliangaugetheory}) the precise definition of this theory requires the choice of the global structure, which can be fixed declaring which of the transformations $A_i\rightarrow A_i+\lambda _i$ are gauge transformations, or, equivalently, specifying the spectrum of line operators. However here the choice is dictated by the global structure of the YM theory. Indeed in the Chevalley basis the eigenvalues of $h_i$ on the weight states of any representations are the Dynkin labels, which must be integer numbers. These are precisely the charges of the Abelian Wilson lines written for the connection $A_i$ in this basis. Therefore the global structure of the $U(1)^{N-1}$ theory we need is that in which, when the Killing form is the Cartan matrix, all the Wilson lines have integer charges.

At the global level this Abelian theory cannot be the correct UV description of YM theory, since it has a $S_N$ 0-form global symmetry, which is instead gauged in YM theory. The action of the permutation group on the $N-1$ field strengths is more evident in the basis defined by the quadratic form $Q^{(N-1)}_{ij}$ defined in (\ref{eq:definitionQ}), which we dub \emph{symmetric basis}. Therefore it is worth to pause a bit to discuss the relation between the two basis of interests. We look for a matrix $L$ such that
\begin{equation}
    A_i=L_{ij}\mathcal{A}_j \ \ \ \Rightarrow \ \ \ L^TA^{(N-1)}L=Q^{(N-1)}
\end{equation}
Where $A^{(N-1)}$ is the Cartan matrix of $\mathfrak{su}(N)$. We solve this constraint using the Cholesky decomposition for both $A^{(N-1)}$ and $Q^{(N-1)}$, namely $A^{(N-1)}=H^TH$, $Q^{(N-1)}=G^TG$, where $H,G$ are upper triangular matrices. Then $L$ is uniquely defined as $L=H^{-1}G$. It turns out that $L$ is upper triangular, with all non-zero components equal to 1:
\begin{equation}
    L_{ij}=\left\{\begin{array}{cc}
      1   & \mbox{if} \ \ i\leq j  \\
      0   &  \mbox{if} \ \  i>j
    \end{array}\right. \ \ \ \ \Rightarrow \ \ \  A_i = \sum_{j\ge i}\mathcal{A}_j
\end{equation}
Notice that $\text{det}(L)=1$, so $L\in GL_{N-1} (\mathbb{Z})$ is an automorphism of the lattice $\mathbb{Z}^{N-1}$. Thus with the global structure dictated by $SU(N)$ YM theory the charges of the Wilson lines are integers in both the Chevalley and symmetric basis.

Let us now discuss in some detail the local operators. For this discussion the suitable basis is the symmetric one. In YM theory all the local operators are constructed out of the non-Abelian field strength $F$, and we can classify them with the powers of $F$ appearing. In the Abelian theory $U(1)^{N-1}$ all the field strengths $\mathcal{F}_{i=1,...,N-1}$ are gauge invariant, so that this theory has much more local gauge invariant operators, and cannot be the UV description of YM theory. By gauging the discrete 0-form symmetry $S_N$ described above we completely fix this mismatch of operators. Let us see how this goes. A general gauge invariant operator of $SU(N)$ YM theory at finite coupling is a product of single-trace operators involving a generic power $k$ of the field strengths\footnote{When writing powers of $F$ we imagine the contraction being given by either $\wedge$ or $\wedge *$, the following analysis holds for both cases.}. In general however, for finite values of $N$, not all single trace operators are independent as there are complicated trace relations among them\footnote{ The simplest example of such relations is for $N=2$ as, for a generic $2\times 2$ hermitian matrix $F$, we have
\begin{equation}
    \text{Tr}\left(F^{3}\right) = \dfrac{1}{2}\text{Tr}\left(F\right)\left[3\text{Tr}\left(F^{2}\right)-\text{Tr}\left(F\right)^{2}\right]
\end{equation}
and similar expressions may be derived for any $\text{tr}\left(F^{k}\right)$ for $k\ge 3$.}, whose origin is simple to understand. The field strength is $N\times N$ hermitian traceless matrix $F$, and we can diagonalize it by rotating it in the Cartan subalgebra
\begin{equation}
     U^{\dagger}FU = \text{diag}(F_{1},.., F_{N}), \ \ \ \ \ \ \  U^{\dagger} U = 1 \quad \ \ \ \ \sum_{i}F_{i}=0
\end{equation}
where $F_i$ are the field strengths of the maximal torus $U(1)^{N-1}$. The trace of the $k$-th power of $F$ is
\begin{equation}
\text{Tr}(F^{k}) = \sum_{i=1}^{N}F_{i}^{k}
\end{equation}
which is manifestly invariant under the Weyl group $S_N$ acting by permutations of the eigenvalues. Such invariance follows from the trace being well-defined on the singular space $U(1)^{N-1}/S_N$ which labels conjugacy classes. We then regard the single-trace operators in $SU(N)$ YM theory as symmetric polynomials in $N$ variables. Since for $N$ variables there are only $N$ independent elementary symmetric polynomials of degree $k\le N$, all local gauge invariant operators of $SU(N)$ YM theory can be expressed as polynomials in the field strengths of the Cartan subalgebra. Since the matrix $F$ is also traceless we have to impose the constraint $\sum_{i}\lambda_i=0$ which reduces the number of independent and non-zero polynomials to $N-1$. To obtain a basis independent statement we may trade the elementary polynomials with $\text{Tr}(F^{k})$, where $k=2,.., N$, so that all other local operators are sums of products of these basic traces. We already discussed the spectrum of gauge invariant local operators of the $U(1)^{N-1}\rtimes S_N$ gauge theory in \ref{sec:semiAbelian}. In the symmetric basis a generic local gauge invariant operator is a symmetric polynomial in the field strengths $\mathcal{F}_{i}$. Again only $N-1$ are independent. The two sets of symmetric polynomials in $SU(N)$ YM theory or $U(1)^{N-1}\rtimes S_N$ are clearly in bijective correspondence, connected by a change of basis in the Cartan subalgebra between the two. It follows that the spectrum of local operators in $SU(N)$ YM theory and in $U(1)^{N-1}\rtimes S_N$ are in one-to-one correspondence. 
\subsection{Line Operators}
\label{sec:line operators}
Now we discuss the Wilson line operators of the $SU(N)$ YM theory
\begin{equation}
    W_{\mathcal{R}}=Tr_{\mathcal{R}}\mathcal{P}\exp{\left(i\oint _{\gamma} A\right)}
\end{equation}
labeled by an irreducible representation $\mathcal{R}$ of the gauge group $SU(N)$. In the full theory they are charged under the $\mathbb{Z}_N$ 1-form symmetry generated by the GW operators corresponding to conjugacy classes in the center $\mathbb{Z}_N\subset SU(N)$, and their charge is the $N-$ality of the representation $\mathcal{R}$.

We want to analyze the UV limit of the Wilson lines. Following the general philosophy that we have described, all the gauge covariant observables can be mapped to the Cartan torus by performing suitable gauge transformations. The holonomy
$$
hol_{\gamma} [A]=\mathcal{P}\exp{\left(i\oint _{\gamma} A\right)}
$$
indeed transform covariantly under $SU(N)$ gauge transformations. Its trace on an irreducible representation $\mathcal{R}$ gives the Wilson line $W_{\mathcal{R}}[\gamma]$. By decomposing the representation in weight states $|\lambda \rangle$ labeled by their Dynkin labels $(\lambda _1,...,\lambda _{N-1})\in \mathbb{Z}^{N-1}$, the trace simply amounts to summing over these states. Since the off-diagonal components of the connections decouple in the UV limit this sum is particularly simple. We express the connection into the Chevalley basis as
$$
A=A_i h_i \ \  , \ \ \ \ \ h_i =2\frac{\alpha _i ^I H^I}{|\alpha _i|^2} \;\; .
$$
The eigenvalues of the Cartan generators in the Chevalley basis are just the Dynkin labels, thus we get a sum of Abelian Wilson lines of the Cartan torus $U(1)^{N-1}$, with charges given by the Dynkin labels. These combinations are always invariant for the action of the Weyl group $S_N$ and then they correspond to a linear sum of the simple Wilson lines of the $U(1)^{N-1}\rtimes S_N$ gauge theory described in section (\ref{sec:semiAbelian}). In order to define carefully this action we have to consider the symmetric basis $\mathcal{A}_i$, on which $S_N$ acts naturally, and then change basis to the connections $A_i$ in the Chevalley basis, in which the Wilson lines are easily written, using $A_i=L_{ij}\mathcal{A}_j$.

To prove the invariance of such lines under $S_N$ we can adopt another point of view. The Wilson lines coincide formally with the characters of the associated representations
\begin{equation}
    \chi(v) = \text{Tr}_{\mathcal{R}} \prod_{i} v_i^{h_i}
\end{equation}
where the product runs over the Chevalley basis and the fugacities $v_i$ are generically complex variables. The Wilson line in representation $\mathcal{R}$ is given by an expression formally identical to the character where the fugacities have been replaced with the holonomies of the components of the gauge field in the Chevalley basis. This proves that the Wilson lines are always invariant under the Weyl group. Indeed the characters are generally defined as the trace of a generic group element in a given representation, as such they are only sensible to the conjugacy class of the element. In other words characters are complex-valued  functions defined on the set of conjugacy classes which, for $SU(N)$, is given by $U(1)^{N-1}/S_N$. It follows that the characters written as Laurent polynomials in the $N-1$ variables corresponding to a maximal torus of $SU(N)$ must be well defined functions on the quotient space $U(1)^{N-1}/S_N$, thus they must be invariant under $S_N$\footnote{As an aside notice that this point of view on the Wilson lines tells us that they fuse exactly as the associated representations of the group which is what should happen at $g_{YM}=0$.}.

Since this discussion is quite abstract we want to present some concrete examples on how to construct these lines for $SU(2)$ and $SU(3)$ YM theories. The reader convinced by the argument above may wish to skip these examples. 

\paragraph{$\boldsymbol{SU(2)}.$} The irreducible representations of $SU(2)$ are characterized by one positive integer $\lambda \in \mathbb{N}$, the Dynkin label of the highest weight state. The states have Dynkin labels $\lambda, \lambda -2,...,-\lambda +2,-\lambda$. In the $g_{YM}\rightarrow 0$ limit the $SU(2)$ Wilson lines $W_{\lambda} ^{SU(2)}$ decompose into a sum over the weight states of the Wilson lines $W(n)=W^n$ of the Abelian theory $U(1)$. In the Chevalley basis the charges $n$ coincide with the Dynkin labels, and we get
\begin{eqnarray}
\label{eq:SU(2) Wilson lines}
W_{\lambda}^{SU(2)}=\sum _{k=0}^{\lambda} W^{\lambda -2k}
\end{eqnarray}
For $SU(2)$ the Chevalley basis and the symmetric one are the same, and indeed the Wilson lines above are manifestly $S_2=\mathbb{Z}_2$ invariant, being a sum of lines $\mathcal{V}(n)=\mathcal{W}^n+\mathcal{W}^{-n}$.

\paragraph{$\boldsymbol{SU(3).}$}The $SU(3)$ case is richer. The weight states in any irreducible representation are labeled by two Dynkin labels $(n_1,n_2)\in \mathbb{Z}^2$, which are the charges of the Wilson lines
\begin{equation}
    W_1^{n_1}=\exp{\left(in_1\oint _{\gamma} A_1 \right)} \ , \ \ \ \ W_2^{n_2}=\exp{\left(in_2\oint _{\gamma} A_2 \right)}
\end{equation}
of the $U(1)^2$ theory expressed in the Chevalley basis. The relation with the symmetric case is $A_1=\mathcal{A}_1+\mathcal{A}_2$, $A_2=\mathcal{A}_2$, so that $
W_1=\mathcal{W}_1\mathcal{W}_2$, $ W_2=\mathcal{W}_2$ and $\mathcal{W}_1=W_1W_2^{-1}$, $ \mathcal{W}_2=W_2$. The action of $S_3$ on the Wilson lines in the symmetric basis is by simple permutations
$$
(\mathcal{W}_1,\mathcal{W}_2,\mathcal{W}_3)\rightarrow (\mathcal{W}_{\sigma (1)}, \mathcal{W}_{\sigma (2)},\mathcal{W}_{\sigma (3)}) \ , \ \  \ \sigma \in S_3
$$
where we should remember that $\mathcal{W}_1\mathcal{W}_2\mathcal{W}_3=1$.
Consider the UV Wilson line in the fundamental representation, whose weight states are $(1,0), (-1,1), (0,-1)$. The Dynkin labels coincide with the charges $(n_1,n_2)$ of the Wilson lines in the Chevalley basis. Hence we have
\begin{equation}
    W_{(1,0)}^{SU(3)}=W_1+W_1^{-1}W_2+W_2^{-1}.
\end{equation}
We can easily check that this operator is $S_3$ invariant. Notice also that the terms above are all mapped into each other by the Weyl group. Indeed by rewriting the lines in the symmetric basis we have
\begin{equation}
    W_{(1,0)}^{SU(3)}=\mathcal{W}_1^{-1}+\mathcal{W}_2^{-1}+\mathcal{W}_1\mathcal{W}_2=\mathcal{V}(-1,0)=\mathcal{V}(0,-1)=\mathcal{V}(1,1)
\end{equation}
namely a single Wilson line of the $U(1)^2\rtimes S_3$ gauge theory. This property is clearly not true for all the representations of $SU(3)$. 

It is worth considering also the anti-fundamental representation, whose weight states are $(0,1), (1,-1), (-1,0)$. The corresponding Wilson line is 
\begin{equation}
    W_{(0,1)}^{SU(3)}=W_2+W_1W_2^{-1}+W_1^{-1}
\end{equation}
which is again $S_3$ invariant. Notice that we can obtain this Wilson line from the one in the fundamental by acting with
$$
C\cdot W_1= W_2 \ , \ \ \ \ C\cdot W_2=W_1.
$$
The operator $C$ is \emph{charge conjugation}. At the level of the connections it exchanges $A_1\leftrightarrow A_2$, thus leaving the Lagrangian $F_1^2+F_2^2-F_1F_2$ invariant. However, as we have just seen, $C$ can act non-trivially on gauge-invariant operators and therefore it is a global symmetry of the theory. This has to be contrasted with $S_3$ which leaves the action invariant, but acts trivially also on the gauge invariant operators. This is because the Weyl group $S_3$ is gauged in the YM theory, while charge conjugation is a 0-form global symmetry acting as an automorphism of the set of line operators.

\subsection{Gukov-Witten Operators}
\label{sec:surface operators}
The surface operators of YM theory, introduced by Gukov and Witten in \cite{Gukov:2006jk, Gukov:2014gja}, are of two types, electric and magnetic. Both types are labeled by conjugacy classes of the gauge group, namely points in $\boldsymbol{\alpha} \in U(1)^{N-1}/S_N$. The electric GW operators labeled by elements of the center $\mathbb{Z}_N\subset SU(N)$ are topological and generate the 1-form center symmetry $\mathbb{Z}_N^{(1)}$ acting on Wilson lines with charge given by the $N$-ality of the associated representation.
In the semi-Abelian theory we similarly have electric and magnetic surface operators, denoted $\mathcal{T}(\boldsymbol{\alpha})$ and $\widetilde{\mathcal{T}}(\boldsymbol{\alpha})$ respectively. As we have seen these are labeled by $\boldsymbol{\alpha}\in U(1)^{N-1}/S_N$, thus exactly matching those of the $SU(N)$ theory. 

A further confirmation that the surface operators of the semi-Abelian theory are related to those of YM theory comes from the action on Wilson lines. The center symmetry of $SU(N)$ is preserved along the RG flow hence must be present also in the deep ultraviolet and should be realized in the semi-Abelian theory. We have already shown that the largest invertible symmetry inside the 2-category describing the surface operators is $\mathbb{Z}_N^{(1)}$ and that these defects act on simple Wilson lines multiplying them by a phase
\begin{equation}
    \mathfrak{C}( \boldsymbol{\alpha}_k,\boldsymbol{n}) = \exp{\left(\frac{2\pi i k}{N}|\boldsymbol{n}|\right)}.
    \label{eq:N-ality}
\end{equation}
To prove that this $\mathbb{Z}_N$ subgroup of the non-invertible symmetry corresponds to the one-form symmetry of the YM theory we need to check that the $SU(N)$ Wilson lines have definite charge proportional to the N-ality of the representation. Notice that a priory this is not obvious since the lines of $SU(N)$ are combinations of the lines of the semi-Abelian theory, and so for generic GW operator $\mathcal{T}(\boldsymbol{\alpha})$ 
\begin{equation}
    \mathcal{T}(\boldsymbol{\alpha})\mathcal{W}^{SU(N)} \not\propto \mathcal{W}^{SU(N)}.
\end{equation}
Actually this factorization occurs precisely for the GW operators generating the center symmetry $\mathbb{Z}_N$. In order to see this we have to rewrite the charge $|\boldsymbol{n}|$ appearing in \eqref{eq:N-ality} in the Chevalley basis. From $A_i = L_{ij}\mathcal{A}_i$ we get $n_i=L_{ji}q_j$, where $q_j$ are the charges in the Chevalley basis. By noting that $\sum _i L_{ji}=j$ we obtain
\begin{equation}
|\boldsymbol{n}|=\sum_{i,j=1}^{N-1} L_{ji} q_j=\sum _j j q_j =:p
    \label{eq:SU(N)Wilson_phases}
\end{equation}
where $p = \sum_i i q_i \; \mbox{mod }N$ is precisely the $N$-ality of the weight state  $(q_1,...,q_{N-1})$. An $SU(N)$ Wilson line in representation $\mathcal{R}$ is a particular combination of simple $S_N$-invariant lines with charges given by the weights of $\mathcal{R}$. Since each weight of a weight system belongs to the same congruence class all terms in the $SU(N)$ Wilson line have same charge under the $\mathbb{Z}_N$ generators. Thus on $SU(N)$ Wilson lines the action of the invertible GW operators factorizes and assigns a charge exactly coinciding with the $N$-ality of the representation. Notice that we found this action only after implicitly imposing a global structure for the semi-Abelian theory dictated by choosing $SU(N)$ as the gauge group of YM theory. Other choices of global structure will lead to different group-like symmetries, this will be discussed in the next subsection.

 \subsection{Global Structures}
 \label{sec:global_structures}
For a gauge theory with Lie algebra $\mathfrak{g}$ we have different choices of global structures, corresponding to different choices of genuine line operators of the theory \cite{Aharony:2013hda}, which can be related by the gaugings of the center symmetry (or some subgroup of it) \cite{Kapustin:2014gua}. In this section we show that all the possible global structures of $\mathfrak{g}=su(N)$ YM theories are nicely matched in the $U(1)^{N-1}\rtimes S_N$ gauge theory\footnote{A similar idea is used in \cite{DelZotto:2022ras} in order to derive the possible choices of global structures of supersymmetric gauge theories from the infrared the Coulomb branch. Here we perform the somewhat complementary analysis in the ultraviolet.}.
\subsubsection{Dirac quantization condition and 't Hooft lines}
In 4d Maxwell theories the possible global structures are the solutions of the Dirac quantization condition. For a single Abelian gauge field the only compact global structure is $U(1)$ and the usual Dirac quantization condition imposes that the charges $q$ and $\widetilde{q}$ of the Wilson and 't Hooft lines respectively must satisfy the condition $q \widetilde{q} \in \mathbb{Z}$. In the case of $U(1)^{N-1}$ Maxwell theory this condition is a straightforward generalization, if we consider the diagonal action
\begin{equation}
     S=\frac{1}{2}\int d^4 x \widehat{F}_i \wedge * \widehat{F}_i
\end{equation}
we have
\begin{equation}
    q_i \widetilde{q}_i \in \mathbb{Z} \ , \ \ \ \forall i=1,...,N-1.
    \label{eq:diagonalDirac}
\end{equation}
These charges however do not have an immediate interpretation in terms of the relation with $SU(N)$ YM theory. To have such interpretation we should work in the Chevalley basis (or the symmetric one) which is non-diagonal. By changing basis $\widehat{F}_i=R_{ij}F_j$, so that the action in the $A_i$ variables is \eqref{eq:action UV YM}, then $K = R^TR $. By denoting with  $n_i,\widetilde{n}_i$ the electric and magnetic charges in the basis with quadratic form $K$, we get
\begin{equation}
    q_i = n_j(R^{-1})_{ji}\ , \  \;\;\; \widetilde{q}_i = \widetilde{n}_j (R^{-1})_{ji} \; \; .
\end{equation}
The Dirac condition (\ref{eq:diagonalDirac}) can now be written as (not summed over $i$)
\begin{equation}
    q_i \widetilde{q}_i = n_j(R^{-1})_{ji} \widetilde{n}_k (R^{-1})_{ki} \in \mathbb{Z}  .
   \end{equation}
Then by summing over $i$ we get
\begin{equation}
 \label{eq:Dirac_condition}
    n_i (K^{-1})_{ij} \widetilde{n}_j \in \mathbb{Z}.
\end{equation}
A particular choice of the global structure in the $U(1)^{N-1}\rtimes S_N$ gauge theory will constraint the set of possible $n_i$, or equivalently the set of possible $\widetilde{n}_i$. Then the constraints on the other charges are completely fixed by \eqref{eq:Dirac_condition}.
The 't Hooft lines of the $U(1)^{N-1}\rtimes S_N$ gauge theory are of the form
\begin{equation}
    \mathcal{M}(\boldsymbol{\widetilde{n}}) = \sum_{\sigma\in S_N} \widetilde{\mathcal{W}}(\mathfrak{S}_{\sigma} ^{\vee}\cdot \widetilde{\boldsymbol{n}}).
\end{equation}
The UV limit of the $SU(N)$ 't Hooft lines are particular combinations of the $\mathcal{M}(\widetilde{\boldsymbol{n}})$ for various $\widetilde{\boldsymbol{n}}\in \mathbb{Z}^{N-1}$ such that the quantity
\begin{equation}
 |\widetilde{\boldsymbol{n}}|=   \sum_{i=1}^{N-1}\widetilde{n}_i
\end{equation}
is fixed. As for the Wilson lines $|\widetilde{\boldsymbol{n}}|$ is the $N$-ality of the corresponding $SU(N)$ representation.
By keeping this in mind we are ready to discuss the relation between the possible global structures of YM theory and those of the semi-Abelian theory.

\subsubsection{Matching the global structures}
To match the $SU(N)$ global structure in the $U(1)^{N-1}\rtimes S_N$ theory we require the charges $n_i$ of the Wilson lines in the Chevalley basis to be all possible integers. With this choice all the UV Wilson lines defined in section (\ref{sec:line operators}) are genuine line operators of the theory. Taking $K=Q$ in (\ref{eq:Dirac_condition}), and choosing only one $n_i$ different than zero and equal to one we get the constraint
\begin{equation}
    \widetilde{n}_i = Q_{ij}v_j \ , \ \ \ \  v_i\in\mathbb{Z}
    \label{eq:Dirac_conditionQ}
\end{equation}
for the charges of the 't Hooft line $\mathcal{H}(\widetilde{n}_1,\cdots,\widetilde{n}_{N-1})$. The condition (\ref{eq:Dirac_conditionQ}) implies that 
\begin{equation}
|\widetilde{\boldsymbol{n}}|=    \sum_{i=1}^{N-1 }\widetilde{n}_i = \sum_{i,j=1}^{N-1 }Q_{ij}v_j = N \sum_{i}v_i \in N\mathbb{Z}
 \end{equation}
where we used $\sum _j Q_{ij}=N$. As expected only the 't Hooft lines with $0$ $N$-ality are genuine line operators. Notice that in this case the invertible magnetic GW operators do not have charged operators, hence only the electric $\mathbb{Z}^{(1)}_N$ is non trivial. By exchanging the roles of $\boldsymbol{n}$ and $\boldsymbol{\Tilde{n}}$ we immediately see that also the global structure of $PSU(N)$ can be reproduced in the semi-Abelian theory, in this case the electric $\mathbb{Z}^{(1)}_N$ invertible symmetry has no charged operator and the one-form symmetry $\mathbb{Z}_{N}^{(1)}$ of the theory is entirely generated by the invertible magnetic GW operators.

The $SU(N)$ and $PSU(N)$ theories are connected by the gauging of the center symmetry. We want to show that also in the UV theory the same conclusion is true. Indeed in the previous section we have shown that $U(1)^{N-1}\rtimes S_N$ posses a $\mathbb{Z}_N$ 1-form symmetry which can be gauged. The action of this group on the Wilson lines of the theory is presented in (\ref{sec:surface operators}) and it is
\begin{equation}
    \mathcal{T}(\boldsymbol{\alpha}_k)\cdot \mathcal{V}(\boldsymbol{n}) = e^{\frac{2\pi i k}{N}|\boldsymbol{n}|} \mathcal{V}(\boldsymbol{n}).
\end{equation}
After gauging only the Wilson lines satisfying $|\boldsymbol{n}|= 0 \text{ mod } N$ remain as good operators of the theory, matching the spectrum of Wilson lines in the $PSU(N)$ theory\footnote{We have also a different but equivalent way to gauge this symmetry. Indeed the GW operators generating $\mathbb{Z}_N$ are a subgroup of the $(U(1)^{(1)}_e)^{N-1}$ symmetry of the $U(1)^{N-1}$ gauge theory before the $S_N$ gauging. Then we can gauge this subgroup in this theory and then gauge the permutation symmetry in the resulting theory. As known, gauging a $\mathbb{Z}_N$ symmetry in a Maxwell theory simply changes the quantization conditions for the electric and magnetic charges and we can easily get the same result obtained in the main text.}. Moreover since now we have eliminated some Wilson lines in the theory, the Dirac quantization condition for the genuine 't Hooft lines
\begin{equation}
    n_i (Q^{-1})_{ij}\widetilde{n}_j \in \mathbb{Z}
\end{equation}
implies that 
\begin{equation}
    n_i \in \frac{1}{N} Q_{ij} v_j \;\;(v_i \in \mathbb{Z})
\end{equation}
which imposes that $|\widetilde{\boldsymbol{n}}| \in \mathbb{Z}$ as it should in $PSU(N)$. It is straightforward to check that gauging $\mathbb{Z}_{l}$ subgroups of the center symmetry one gets a spectrum of lines in the semi-Abelian theory which exactly matches the spectrum of the $SU(N)/\mathbb{Z}_l$ gauge theory. 

\section{Outlook}
\label{sec:4}
The main motivation of this paper was studying the properties of the continuous non-invertible symmetries arising in the $U(1)^{N-1} \rtimes S_N$ gauge theories and make a connection with the UV limit of $SU(N)$ YM theory. In particular we have found that all the GW operators of the non-Abelian theories become topological in the deep UV and they describe a non-invertible symmetry which is broken to its group-like subcategory $\mathbb{Z}_N$ along the RG flow. Therefore this is one of the few examples in which the gauging of an automorphism is not an artificial mechanism introduced to produce non-invertible symmetries but instead comes naturally from physically interesting systems. In doing this we have analyzed extensively the symmetry, which forms a continuous 2-category with an intricate structure arising form the presence of topological lines, appearing as 1-morphisms. The fusion rules encodes information about these morphisms in the integer constants $f_{ab}^{\sigma}$, and in the presence of the condensation defects.

Even if we analyzed explicitly the $SU(N)$ gauge theory, it is easy to see that our results extend to any gauge group $G$. The theory encoding the gauge invariant data in the ultraviolet is the $U(1)^r\rtimes W_G$ gauge theory, where $r$ is the rank of $\mathfrak{g}=\mbox{Lie}\;G$ and $W_G$ is the Weyl group. Then the fusion rules \eqref{eq:local fusions} as well as the action of the GW on line operators \eqref{eq:actionWilson} are simply obtained by replacing $S_N$ with $W_G$. Also the analysis of the condensation defects, the global fusions and the 2-categorical structure is conceptually identical for any gauge group $G$.

We conclude by proposing interesting open problems which arise naturally from our work, and also give qualitative ideas and suggestions about these issues.

\textbf{Non-local currents, spontaneous symmetry breaking and anomalies.}
The first question concerns the properties of the continuous non-invertible symmetries studied in this paper. Indeed it is natural to ask if such symmetries have conserved currents and if possible spontaneous symmetry breaking of continuous non-invertible symmetry would lead to Goldstone bosons. The existence of conserved currents can be derived from the known conserved currents of the $U(1)^{N-1}$ theory before the $S_N$ gauging. In this theory we have the conserved $2$-form current
\begin{equation}
    j^i = F^i
\end{equation}
where $i = 1,\cdots,N-1$, corresponding to the $(U(1)^{(1)})^{N-1}$ 1-form symmetry of the theory. 
\\After the $S_N$ gauging this operator is no longer gauge invariant and then it cannot be regarded as a good operator of the theory. However we can construct a gauge invariant non-genuine local operator attaching to $F^i$ an $S_N$ Wilson line in the $N-1$ standard representation 
\begin{equation}
    J = W_{S_N}(\gamma_x)^{i} F^i(x).
\end{equation}
In the above equation $\gamma_x$ is an infinite topological line which ends on $x$ and then $J$ is a good gauge invariant operator. The idea is that currents of non-invertible symmetries correspond to non-genuine local operators \cite{Thorngren:2021yso}. Note that however this new current is not conserved but is covariantly conserved with respect to $S_N$ transformations, namely
\begin{equation}
    D_{S_N} J = 0.
\end{equation}
In particular the conserved current in ordinary invertible symmetries is the operator creating Goldstone particles from the vacuum when such a symmetry is spontaneously broken. In this case it would be interesting to understand what happens to these excitations and interpreting them from a generalized version of a Goldstone theorem\footnote{For a generalization of the Goldstone theorem in the case of \emph{invertible} higher-form symmetries see e.g. \cite{Lake:2018dqm,Hofman:2018lfz}.}.   

Another interesting question is about the possible mixed 't Hooft anomaly between the electric and magnetic non-invertible symmetries possessed by the semi-Abelian gauge theory. Indeed before the $S_N$ gauging the $U(1)^{N-1}$ gauge theory has such an anomaly between the invertible 1-form symmetries $(U(1)_e^{(1)})^{N-1}$ and $(U(1)_m^{(1)})^{N-1}$. This anomaly involves continuous symmetries and we expect it to be inherited by the non-invertible symmetries since a discrete gauging cannot cancel a continuous anomaly. However to study this anomaly we need to couple these symmetries to backgrounds (note that the $B_{e,m}$ backgrounds of the Abelian theory are not anymore gauge invariant) but a consistent definition of backgrounds for non-invertible symmetries is still an open problem.

\textbf{Constraints on the RG flow of Yang-Mills theories.}
Perhaps the most important question regards possible implications of the UV emergent symmetries along the RG flow of YM theories. Indeed in a generic QFT, a symmetry possessed by the UV fixed point and broken by some relevant deformations affects the possible structure of the low energy effective theory. This is the case, for instance, of the quark mass perturbation in QCD which leads to mass terms in the chiral Lagrangian. In this case it would be interesting to study more carefully the deformation which breaks this non-invertible symmetry to the center symmetry of YM theory. In particular we expect that for instance correlation functions involving a GW operator and a Wilson line
\begin{equation}
    \langle T(\boldsymbol{\alpha})^{SU(N)} [\Sigma_2] W_{R}(\boldsymbol{n})^{SU(N)}[\gamma]...\rangle,
\end{equation}
 which at $g\not=0$ and $T(\alpha) \not\in \mathbb{Z}_N$  depends on the relative position of the surface $\Sigma_2$ and the curve $\gamma$, when the surface is infinitesimally closed to $\gamma$, they approximately follow the topological action presented in the previous sections, with corrections of order $\Lambda_{YM}r$ where $r$ parametrizes the distance between $\Sigma_2$ and $\gamma$\footnote{For instance taking $\Sigma_2 = S^2$ surrounding $\gamma$ then $r$ is exaclty the radius of the sphere. }. 
 \\We hope that other possible predictions can be achieved also when the issues presented in the first part of this section will be understood. In particular the presence of an anomaly before the deformation would suggest that the gap produced by the RG flow should go to zero in the limit in which the RG flow is never triggered. Indeed this is something believed to happen in YM theory since the gap is of order $\Lambda_{YM}$.
 \section*{Acknowledgments}

We would like to thank Stephane Bajeot, Francesco Benini, Christian Copetti, Thibault Decoppet, Lorenzo Di Pietro, Marco Serone and Matthew Yu for useful discussions. We especially thank Christian Copetti and Marco Serone for useful comments on the draft which led us to a better understanding of some of the results presented in this work. We are grateful to the Perimeter Institute for Theoretical physics for the hospitality and to the organizers of the workshop "Global Categorical Symmetries", during which this work has been completed and presented in a poster session. Research at Perimeter Institute is supported in part by the Government of Canada through the Department of Innovation, Science and Economic Development Canada and by the Province of Ontario through the Ministry of Colleges and Universities. This work is partially supported by INFN Iniziativa Specifica ST\&FI. A.A. and G.R. are supported in part by the ERC-COG grant NP-QFT No. 864583 ``Non-perturbative dynamics of quantum fields: from new deconfined phases of matter to quantum black holes''.
 
\appendix
\section{The Permutation Group $S_N$}
\label{sec:SN}
In this appendix we describe some representations of $S_N$ used in the main text. The simplest possible action is the natural representation, which consists of the usual permutations of $N$ variables. We collect the variables $\mathcal{F}_i$ in a vector $\boldsymbol{\mathcal{F}}$ and denote the action of $\sigma \in S_N$ as $\sigma \cdot \boldsymbol{\mathcal{F}} = (\mathcal{F}_{\sigma(i)},..,\mathcal{F}_{\sigma(N)})$. Clearly this representation is reducible. The vectors with all equal entries are fixed by all permutations and span the one-dimensional trivial representation. The orthogonal complement of this subspace is given by those vectors whose components sum to zero. Thus we may construct an $N-1$ dimensional irreducible representation imposing the $S_{N}$-invariant constraint
\begin{equation}    \sum_{i=1}^{N}\mathcal{F}_i =0.
\end{equation}
This defines the standard representation, of dimension $N-1$, which we denote as $\mathfrak{S}$. We construct the dual $N-1$ dimensional representation as follows. Let us introduce another set of $N$ variables $u_i$, collected in a vector $\boldsymbol{u}$, and consider the scalar product $\boldsymbol{u}\cdot \boldsymbol{\mathcal{F}} = \sum_{i=1}^{N}u_i \mathcal{F}_i$.
We define the representation dual to the one carried by $\boldsymbol{\mathcal{F}}$ as the representation on $\boldsymbol{u}$ which preserves the scalar product. This means, for a pair of dual representations $R^{\vee}$ and $R$ acting on vectors $\boldsymbol{a}$ and $\boldsymbol{b}$ respectively
\begin{equation}
    R^{\vee}(\boldsymbol{a})\cdot R(\boldsymbol{b}) = \boldsymbol{a}\cdot \boldsymbol{b} \ .
\end{equation}
Clearly acting with a permutation on the $u_i$ or the $\mathcal{F}_i$ is equivalent, in this sense the natural representation is self-dual. For the standard representation, solving the constraint for $\mathcal{F}_N$, we get
\begin{equation}
    \sum_{i=1}^{N}u_i \mathcal{\mathcal{F}}_i= \sum_{i=1}^{N-1}(u_i-u_N) \mathcal{\mathcal{F}}_i = \sum_{i=1}^{N-1}\alpha_i \mathcal{\mathcal{F}}_i  \ .
\end{equation}
The $N-1$ coefficients $\alpha_i = u_i - u_N$ carry the representation $\mathfrak{S}^{\vee}$, dual to $\mathfrak{S}$, defined as
\begin{equation}
    \sum_{i,k}\alpha_{i} (\mathfrak{S}_{\sigma})_{ik}\mathcal{F}_{k} = \sum_{i,j} (\mathfrak{S}^{\vee}_{\sigma})^{-1}_{ij}\alpha_{j} \mathcal{F}_{i} \rightarrow  \mathfrak{S}^{\vee}_{\sigma} = (\mathfrak{S}_{\sigma^{-1}})^{T} \ .
\end{equation}
Thus the explicit action of $\sigma \in S_N$ is
\begin{equation}
    \mathfrak{S}^{\vee}_{\sigma}(\alpha_i) = u_{\sigma(i)} - u_{\sigma(N)} \ .
\end{equation}
This representation respects the composition covariantly:
\begin{equation}
    \mathfrak{S}_{\sigma _1}^{\vee}\circ \mathfrak{S}_{\sigma _2}^{\vee}=\mathfrak{S}_{\sigma _1\circ \sigma _2}^{\vee} \ .
\end{equation}
In the symmetric basis used in the text the action on the gauge fields is given by the standard representation. Consequently the action on both the charges of the Wilson lines and the continuous parameters of the GW operators is given by the dual representation $\mathfrak{S}^{\vee}$.
\section{Some detail on pure $\text{Rep}(G)$ gauge theories}
\label{sec:2dTQFT}
As we discussed in the main text, the fusion coefficients of the condensation defects $\cC_{\text{Rep}(S_N)}$ involve the partition function of the pure $\text{Rep}(S_N)$ gauge theory. Here we will provide some details on the characterization of the pure $\text{Rep}(G)$ gauge theories for any finite group $G$, in terms commutative Frobenius algebras.

TQFTs in 2d are particularly simple because every space-like slice is a disjoint union of circles, and any compact surface $\Sigma$ can be constructed by gluing pair of pants. Because of the first fact the only Hilbert space we need to assign is $\cH _{S^1}$, while the second implies the well known result that 2d TQFTs are fully determined by commutative Frobenius algebra structure on this Hilbert space \cite{Abrams}. Formally this is obtained by specifying an associative and commutative multiplication $\mu : \cH _{S^1}\otimes \cH _{S^1}\rightarrow \cH _{S^1}$ and a linear map $\theta : \cH _{S^1}\rightarrow \bC$. More concretely, the Hilbert space inherits an algebra structure from the one of local operators, by using operator/state correspondence, while $\theta$ acts by taking the scalar product with the Hartle-Hawking state corresponding to the identity operator.

As a simple example, which is then easy to generalize to the case of $\text{Rep}(G)$, consider the pure $\bZ _N$ gauge theory in 2d. A clear construction is by starting from the trivial theory with $\bZ _N$ symmetry, namely a theory of $N$ line operators fusing according to $\bZ_N$, and no local operator. Thus the Hilbert space is trivial, and the symmetry does not act on anything. Each line, however, has a non-empty twisted sector containing one operator. After gauging, these twist operators become local, are labeled by elements of $\bZ _N$ and they fuse accordingly. There are also new line operators generating the dual $\widehat{\bZ}_N=\text{Hom}(\bZ _N, U(1))$ symmetry. They are labeled by irreducible representations and act on local operators. By operators/state correspondence the Hilbert space is $N$ dimensional, and it is in a (reducible) representation of the dual $\widehat{\bZ}_N$ symmetry, given by the direct sum of all the irreducible representations, namely the regular representation. This has a clear interpretation in the context of gauging in fusion categories, which can be easily generalized. One can think the gauged $\bZ _N$ symmetry as the category of representations of $\widehat{\bZ}_N$, and the gauging is understood as the insertion of a mesh of the Frobenius algebra objects corresponding to the regular representation of $\widehat{\bZ}_N$ \cite{Bhardwaj:2017xup}. The Hilbert space after gauging is the twisted sector of this algebra object, which therefore forms the regular representation of the dual symmetry. The commutative Frobenius algebra structure of the Hilbert space is then given by the Frobenius algebra structure of the regular representation of $\widehat{\bZ}_N$ and the Hartle-Hawking state corresponds to the singlet representation.

The generalization to the pure $\text{Rep}(G)$ gauge theory is straightforward. We start from the trivial theory enriched with topological lines forming the category $\text{Rep}(G)$. Then we insert a fine mesh of the algebra object in the regular representation of $G$, and the Hilbert space of the gauged theory will be organized in such representation. This naturally has a commutative Frobenius algebra structure, which can be used to define axiomatically the theory. The dual symmetry is now generated by lines fusing according to the $G$ group law, so that it is an invertible symmetry, possibly non-abelian.

\newpage
\bibliographystyle{ytphys}
\bibliography{paper}

\end{document}